\documentclass[useAMS,usegraphicx,usenatbib]{mn2e}
\usepackage{rotating,graphicx,subfig}

\def\ud{{\mathrm d}}
\def\eg{{\it e.g.\ }}
\def\degrees{^\circ}
\newcommand{\greeksym}[1]{{\usefont{U}{psy}{m}{n}#1}}

\newcommand{\uDelta}{\mbox{\greeksym{D}}}

\newcommand{\lt}{\mathrel{\raise2pt\hbox{\rlap{\hbox{\lower6pt\hbox{$\sim$}}}\hbox{$<$}}}}
\newcommand{\gt}{\mathrel{\raise2pt\hbox{\rlap{\hbox{\lower6pt\hbox{$\sim$}}}\hbox{$>$}}}}

\defcitealias{dl07}{DL07}
\defcitealias{dl08}{DL08}
\defcitealias{dl09}{DL09}
\defcitealias{mg11}{MG12}
\defcitealias{napolitano09}{N09}
\defcitealias{foster11}{F11}

\title[Dark matter halo of NGC~4494]{Elliptical galaxies with rapidly
  decreasing velocity dispersion profiles: NMAGIC models and dark halo
  parameter estimates for NGC~4494} 
\author[L. Morganti et al.]{Lucia Morganti$^{1}$\thanks{E-mail: morganti@mpe.mpg.de},
Ortwin Gerhard$^1$\thanks{E-mail: gerhard@mpe.mpg.de}, Lodovico Coccato$^2$, Inma Martinez-Valpuesta$^1$, 
\and Magda Arnaboldi$^{2,3}$
\\$^1$ Max-Planck-Institut f\"ur extraterrestrische Physik, Postfach 1312, 
Giessenbachstr., D-85741 Garching, Germany
\\$^2$ European Southern Observatory, Karl-Schwarzschild-Straße 2, D-85748 Garching, Germany
\\$^3$ INAF, Osservatorio Astronomico di Pino Torinese,
          I-10025 Pino Torinese, Italy}

\begin{document}
\date{MNRAS accepted.}

\maketitle

\begin{abstract}   

  NGC~4494 is one of several intermediate-luminosity elliptical
  galaxies inferred to have an unusually diffuse dark matter halo.  We
  use the $\chi^2$-made-to-measure particle code NMAGIC to construct
  axisymmetric models of NGC~4494 from photometric and various
  kinematic data. The extended kinematics include light spectra in
  multiple slitlets out to $3.5 R_{\rm e}$, and hundreds of planetary
  nebulae velocities out to $\simeq 7 R_{\rm e}$, thus allowing us to
  probe the dark matter content and orbital structure in the halo.

  We use Monte Carlo simulations to estimate confidence boundaries for
  the halo parameters, given our data and modelling set-up. We
    find that the true potential of the dark matter halo is recovered
    within $\Delta G$({\rm merit function})$\lt26$
    ($\Delta\chi^2\lt59$) at 70\% confidence level (C.L.), and within
    $\Delta G\lt32$ ($\Delta\chi^2\lt70$) at 90\% C.L.. These numbers
  are much larger than the usually assumed $\Delta\chi^2=2.3 (4.6)$
  for 70\% (90\%) C.L.  for two free parameters, perhaps
  case-dependent, but calling into question the general validity of
  the standard assumptions used for halo and black hole mass
  determinations.

  The best-fitting models for NGC~4494 have a dark matter fraction of
  about $0.6\pm0.1$ at $5R_{\rm e}$ (70\% C.L.), and are embedded in a
  dark matter halo with circular velocity $\sim200{\rm km s}^{-1}$.
  The total circular velocity curve (CVC) is approximately flat at
  $v_{\rm c}=220 {\rm km s}^{-1}$ outside $\sim0.5R_{\rm e}$. The
  orbital anisotropy of the stars is moderately radial.  These results
  are independent of the assumed inclination of the galaxy, and
  edge-on models are preferred.  Comparing with the halos of NGC~3379
  and NGC~4697, whose velocity dispersion profiles also decrease
  rapidly from the center outwards, the outer CVCs and dark
    matter halos are quite similar. NGC~4494 shows a particularly
  high dark matter fraction inside $\sim3R_e$, and a strong
  concentration of baryons in the center.

\end{abstract}   


\section{Introduction}

The formation and evolution of elliptical galaxies has been an
important issue in extragalactic astrophysics since a long time.
Being collisionless to a very good approximation, ellipticals retain
relics of their formation history in the present-day orbital
structure, especially in their halos due to the longer dynamical time
scales.  According to the currently favoured hierarchical formation
scenario in a $\Lambda$CDM cosmology, these halos are dark matter
dominated.  The ambitious task of inferring both the orbital structure
and mass distribution of ellipticals is commonly tackled by dynamical
modelling of the observational data.

Unfortunately, the lack of an ubiquitous tracer such as HI gas in
spiral galaxies \citep[but see][]{bertola93,franx94,oosterloo02} makes
mass measurements in elliptical galaxies challenging.  The strongest
evidence for dark matter halos is found for giant ellipticals whose
mass distribution can be determined from X-ray emission of the hot gas
\citep[\eg][]{loewenstein99,humphrey06,das10} or strong gravitational
lensing techniques \citep[\eg][]{maoz93,keeton01,treu04,auger10}.
These studies are consistent with massive dark halos, and nearly
isothermal total mass profiles \citep{2comp09}.

By contrast, the situation with less massive, X-ray faint ellipticals
is more controversial.  However, dynamical models, particularly when
fitting higher order moments of the line-of-sight velocity
distribution (LOSVD) from stellar absorption lines, eventually
ascertained the presence of dark matter halos around these
intermediate-luminosity ellipticals and are generally consistent with
flat circular velocity curves.  With LOSVD measurements limited by the
rapid fall-off of the stellar surface brightness, the kinematics of
discrete tracers such as planetary nebulae (PNe) and globular clusters
(GCs) usually represent the only possibility to probe the mass
distribution and orbital structure beyond $2R_{\rm{e}}$, in the realm
of dark matter
\citep[\eg][]{hui95,mendez01,pns,peng04,dl08,dl09,napolitano09,payel2,napolitano11,deason12}.

Curiously, the PNe velocity dispersion profiles of some of the nearby
intermediate-luminosity ellipticals show a strong, quasi-Keplerian decline with radius outside $1 R_e$
\citep{coccato09}, suggesting very little (if any) dark matter
\citep{rom03}.  The aim of the present paper is to expand the sample
of modelled quasi-Keplerian  intermediate luminosity ellipticals,
focussing on NGC~4494, and then compare the results with those
previously obtained for other galaxies of this class, namely NGC~4697
\citep[hereafter \citetalias{dl08}]{dl08} and NGC~3379
\citep[hereafter \citetalias{dl09}]{dl09}.

NGC~4494 is an E1-E2 elliptical galaxy in the outer regions of the
Virgo cluster, with a smooth light profile and an intermediate stellar
mass of about $10^{11}M_{\sun}$ \citep[hereafter
\citetalias{foster11}]{foster11}.  It has been variously described as
a loose group member \citep{forbes96} or isolated galaxy
\citep{lackner10}. NGC~4494 is classified as a fast rotator
  by \citet[][$\lambda_{\rm R}\simeq 0.2$]{emsellem11}. Among the
peculiarities of this galaxy, a sharp central ring of dust
\citep{forbes95,lauer05}, a kinematically decoupled core
\citep{bender94,davor11}, moderate rotation $\sim 60{\rm km s}^{-1}$
out to $\sim 3R_{\rm e}$ \citep{proctor09}, and an
  outward-decreasing $\lambda_{\rm R}$-profile \citep{coccato09} have
been reported.  The velocity dispersion profile of NGC~4494 decreases
rapidly from about $160{\rm km s}^{-1}$ in the center to about $70{\rm
  km s}^{-1}$ at $\sim7R_{\rm e}$ \citep[hereafter
\citetalias{napolitano09}]{napolitano09}, hinting to a possible
deficiency in dark matter.  Also, the X-ray flux of NGC~4494 is two
hundred times fainter than that of other galaxies of the same optical
luminosity, which has been interpreted as the result of a recent
interaction which depleted the gas, or again as evidence for little
dark matter \citep{osullivan04,fukazawa06}.

NGC~4494 has been recently modelled using spherical Jeans models
\citepalias{napolitano09}, and axisymmetric particle models
constructed with the iterative method \citep{rod11}.  The available
observational data, which included PNe velocities out to $\sim7R_{\rm
  e}$ \citepalias{napolitano09}, were best fit by low concentration
dark halos, with some uncertainties related to the adopted modelling
assumptions, the differences between the fits for different models
\citepalias{napolitano09}, and the limited number of explored models
\citep{rod11}.

Recently, new observational data consisting of stellar absorption line
kinematics in multiple slitlets out to $\sim3.5R_{\rm e}$ became
available for this galaxy \citep[\citetalias{foster11}]{proctor09}.
Moreover, in the work of \citet{deason12} NGC~4494 appeared as an
outlier with curiously low dark matter fraction within $5R_{\rm e}$,
with respect to model predictions assuming either a Salpeter or a
Chabrier initial mass function.  These facts prompted us to undertake
a further careful analysis of the dark matter content and orbital
structure of NGC~4494, incorporating as many observational data as
currently available and assessing the uncertainties in the recovery of
dark halo parameters via dynamical models.

In this paper, we construct new dynamical models fitting all available
photometric and kinematic data with the flexible particle code NMAGIC
\citep[hereafter \citetalias{dl07}]{dl07}, which implements a modified
version of the made-to-measure (M2M) technique proposed by
\citet{st96}, suitable for the modelling of observational data with
errors ($\chi^2$M2M).  NMAGIC works by slowly correcting the particle
weights of an evolving N-body system until a satisfactory compromise
is achieved between the goodness-of-fit to the observational data, and
some degree of regularization of the underlying particle model.  More
recent implementations of the method can be found in \citet{dehnen09},
who proposed a different technique for the weight adaptation,
\citet{longmao10}, and \citet[hereafter \citetalias{mg11}]{mg11}, who
introduced a Moving Prior Regularization method to generate smooth
$\chi^2$M2M particle models fitting noisy data without erasing the
global phase-space structures.

So far, the M2M technique has been used to investigate the dynamics of
the Milky Way's bulge and disk \citep{bissantz04,long13}, the mass
distribution and orbital structure in the outer halos of elliptical
galaxies \citep[\citetalias{dl08,dl09};][]{payel2}, and the dynamics
of a sample of SAURON elliptical and lenticular galaxies
\citep{longmao12}.  In this paper, we construct axisymmetric NMAGIC
models for NGC~4494 which include PNe \citepalias{napolitano09} and
new stellar absorption line kinematics in slitlets
\citepalias{foster11}, for different dark matter contributions to the
total gravitational potential, and for different inclinations.

One of the key points in this work is to study the accuracy with which
the parameters of our models can be estimated given the observational
data at hand. M2M particle methods work by adjusting the weights
  of a large number of particles in order to achieve a good match to
  the data. Similarly, in the more common Schwarzschild methods a
  large number of orbital weights is adjusted to fit the
  data. Typically, there are many more weights than data constraints,
hence the number of degrees of freedom\footnote{ The number of degrees
  of freedom is defined as the number of constraints (data points plus
  constraints introduced by \eg regularization) minus the number of
  free parameters (model parameters plus fitted weights).}  is likely
to be much smaller than the number of data points
\citep{cretton00,gebhardt00}. However, to determine the
  effective number of free parameters involved in the modelling is
  very difficult.

 It is common practise to use the relative differences
  $\Delta\chi^2=\chi^2-\chi^2_{min}$ to measure confidence limits for
  the subset of parameters specifying the mass model. On the
assumption that the observational errors are Gaussian distributed
\citep[but see][]{vandermarel00}, that the model is linear in the
parameters, and the number of degrees of freedom is positive,
$\Delta\chi^2$ follows $\chi^2$-statistics \citep[\eg][]{press92}.
Many dynamical modelling studies have therefore estimated, e.g., 68\%
confidence limits for 1 (2,3) free model parameter(s) from
$\Delta\chi^2=1 (2.3,3.5)$ contours in these parameters
\citep[\eg][]{vandermarel98,cretton99,barth01,verolme02,cappellari02,gebhardt03,valluri04,shapiro06,chaname08,cappellari09_2,vandenbosch10,murphy11,adams12},
but see \citet{vandenbosch+vandeven09}.  Here, prompted by large
$\Delta\chi^2$ values between quite similar models, we use a Monte
Carlo simulation approach to estimate confidence intervals for the
parameters based on models of NGC~4494-like mock galaxies \citep[see
also][]{press92,gerhard98,thomas05}.

The paper is organized as follows. In Section \ref{sec:data} we
describe the observational data used for NGC~4494.  In
Section~\ref{sec:mod} we outline the modelling technique.  In
Section~\ref{sec:test} we construct NGC~4494-like target galaxies and
their observables in different dark matter halos, and use them to
calibrate the optimal amount of regularization, and to estimate the
confidence levels for parameter estimation with our NMAGIC models.
Dynamical models of NGC~4494 for a range of dark matter halo
potentials and inclinations are then constructed in
Section~\ref{sec:models}.  The main implications of our findings are
discussed in Section~\ref{sec:discuss}.  Finally, the paper closes
with our conclusions in Section~\ref{sec:conclusions}.

\section{OBSERVATIONAL DATA}\label{sec:data}
In this Section we describe the observational data that we will use
for modelling the elliptical galaxy NGC~4494.  We adopt a distance of
15.8~Mpc \citep{tonry01}, so that 1 kpc=$13''$, a systemic velocity
$v=1344{\rm km s}^{-1}$, from NED, and $R_{\rm e}=49''\approx3.77$ kpc
\citep{deV91}.

\subsection{Photometric data and deprojection}
\label{ssec:photdata}
\begin{figure}
\centering
\includegraphics[angle=-90,width=0.85\hsize]{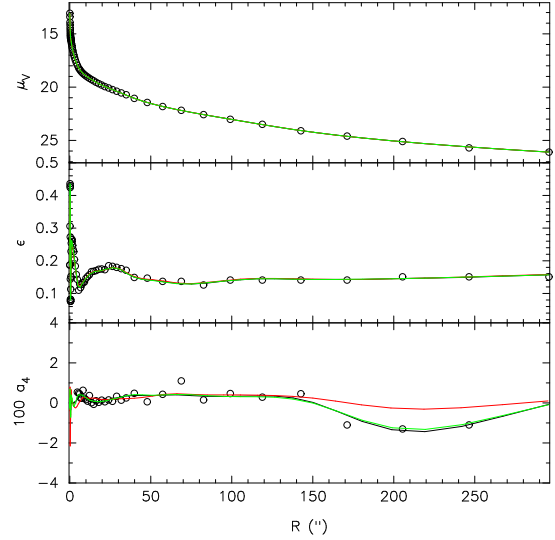}
\caption{{\it From top to bottom}: $V$-band surface-brightness,
  ellipticity, and fourth-order isophotal shape coefficient $a_4$ as a
  function of major-axis projected radius for NGC~4494.  {\it Circles}
  are the measurements, {\it lines} represent the reprojected values
  from axisymmetric deprojections obtained with an assumed inclination
  of 90$\degrees$({\it black}), 70$\degrees$({\it green}), and
  45$\degrees$({\it red}).
 \label{fig:phot}}
\end{figure}

As photometric data, we use the V-band surface-brightness profile,
ellipticity and shape parameter $a_4$ values of
\citetalias{napolitano09}.  The photometric data extend to $273''$
along the major axis, and are a combination of \textit{HST} data in
the V and I bands inside $4.3''$ \citep{lauer05}, ground-based
observations in BVI out to $32''$ \citep{goudfrooij94}, and Megacam
data from the Sloan Digital Sky Survey \textit{$g'$} filter
\citepalias{napolitano09}.  The total extinction-corrected luminosity
in the $V$-band is $2.6\times10^{10} L_{V, \odot}$
\citepalias{napolitano09}.

A S\'ersic fit to the surface brightness profile outside the central
dust region ($R>5.6''$) gives $n=3.30$ \citepalias{napolitano09}.  The
observed ellipticity is $\epsilon=0.15-0.20$ (axis ratio
$q=0.85-0.80$) for $R<R_{\rm e}/2$, outside of which NGC~4494 becomes
rounder, with $\epsilon=0.13-0.15$ ($q=0.87-0.85$) for $1-1.5R_{\rm
  e}$.  The radial profiles of surface-brightness, ellipticity, and
shape parameter $a_4$ are shown in Fig.~\ref{fig:phot}.
 
In our NMAGIC models for the galaxy NGC~4494, we will not use the 2D
surface brightness, but rather its 3D deprojected luminosity density.
The deprojection of the surface brightness is unique only for
spherical or edge-on axisymmetric systems \citep[\eg][]{gerhard96}.
Here, we consider axisymmetric deprojections for inclinations
$i=90\degrees,70\degrees$, and $45\degrees$, which is very close to
the minimum inclination allowed by the observed flattening of NGC~4494
\citepalias{napolitano09,foster11}.  For each inclination angle, we
use the maximum penalized likelihood scheme and program described in
\cite{mag99} to find a smooth axisymmetric density distribution
consistent with the surface-brightness profile.  The method favours a
power-law density profile in the radial region not constrained by
photometric data.

The overall good agreement between the measured and reprojected
surface-brightness, ellipticity, and shape parameter $a_4$ is shown in
Fig.~\ref{fig:phot}. For $i=45\degrees$, the method
  apparently has some difficulties in reproducing the boxy isophotes
  at $R\sim200''$ accurately.

\subsection{Kinematic data}
\label{ssec:kindata}
\begin{figure}
\centering
\includegraphics[angle=-90,width=0.95\hsize]{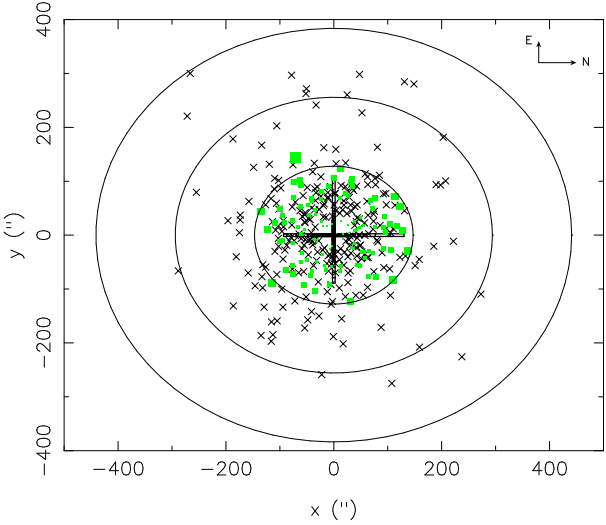}
\caption{Spatial distribution of published kinematic data for
  NGC~4494.  Long-slits along the major and minor axis are plotted,
  where the $x$-axis coincides with the major axis of the galaxy.  The
  positions of PNe are marked by {\it black crosses}, those of the
  slitlets by {\it green squares}.  {\it Ellipses} represent $3,6,$
  and $9R_e$, for an axis ratio $q=0.87$.
\label{fig:setup}}
\end{figure}
We combine three kinematic data sets in order to achieve the widest
possible spatial coverage and probe the mass distribution and orbital
structure far out in the halo of NGC~4494.  In particular, we use
long-slit absorption line kinematics extending out to $\sim2{R_{\rm
    e}}$ \citep{coccato09}, absorption line kinematics in slitlets out
to $\sim3.5{R_{\rm e}}$ \citepalias{foster11}, and PNe line-of-sight
velocities reaching $\sim7{R_{\rm e}}$ \citepalias{napolitano09}.  The
spatial coverage of the combined kinematic constraints can be
appreciated in Fig.~\ref{fig:setup}.

\subsubsection{Stellar-absorption line slit data}
\label{sec:slit}
\begin{figure}
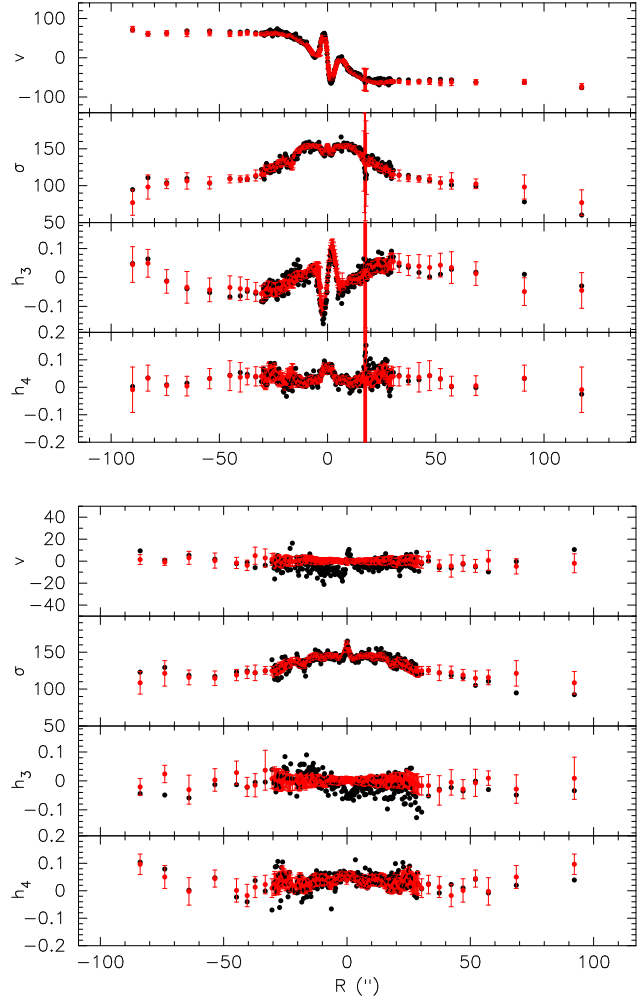

\includegraphics[scale=0.41,angle=-90.0]{4494_maj.ps}
\vskip.5cm
\includegraphics[scale=0.41,angle=-90.0]{4494_min.ps}
\caption[]{Long-slit kinematics along the major ({\it top}) and minor
  axis ({\it bottom}) of NGC~4494.  {\it Black} and {\it red dots}
  represent the original data and the data used for the modelling,
  respectively.  See Section~\ref{sec:slit} for details.}
\label{fig:slit_data}
\end{figure}

Long-slit absorption line kinematics within $\sim 2R_{\rm e}$ were
presented in \citet{coccato09}, and consist of line-of-sight velocity
$v$, velocity dispersion $\sigma$, and higher-order Gauss-Hermite
coefficients $h_3$ and $h_4$, along the major and minor axis of
NGC~4494.

The original data are shown in Fig.~\ref{fig:slit_data} with black
dots.  They are consistent with small or zero rotation along the minor
axis, and substantial major axis rotation, flattening beyond $20''$ at
$V\sim60{\rm km s}^{-1}$ out to $120''$.  In the inner $10''$ along
the major axis, the signature of the decoupled core is apparent.  The
velocity dispersion decreases from about $160{\rm km s}^{-1}$ at the
center to about $80{\rm km s}^{-1}$ at $\sim 100''$.

We noticed a systematic offset between the velocities and $h_3$
coefficients measured along the major and minor axes in the central
arcsec.  This leads us to suspect an offset of the minor axis slit
from the galaxy centre, in the South direction.  The required
offset (less than $1''$) is smaller than the slit width and the
average seeing.  Furthermore, the $h_3$ measurements along the minor
axis are overall negative on both sides, which is unexpected even for
a triaxial system with minor axis rotation.  For the modelling, we
therefore replace the measurements of $v$ and $h_3$ along the minor
axis with Gaussian random variates with zero mean and $1\sigma$
dispersion equal to the observational errors.

Inside $32''$ there are many nearby data points whose respective rms
deviations are larger than their error bars.  In order to reduce their
impact on the modelling, we run a central moving average over the data
within $32''$, averaging over 7 data points (3 points on each side),
and substituting each point with the value of the average.
Because the errors of the neighbouring points are very
  similar, we decided to leave them at their original values.

Then, we minimize the impact of the feature at $\sim20''$ along the
major axis, which is due to contamination from a foreground star, by
artificially increasing the error bars of those measurements.

As we are interested in axisymmetric models of the data, we finally
symmetrize the slit data set, as in \citetalias{dl09}.  In practice,
we average the values of the measured kinematics at two similar radii
($R_+,R_-$) on both sides of the slit with respect to the center.
Since the kinematic data show major axis rotation, we take into
account the sign reversal of $v$ and $h_3$ when symmetrizing the major
axis slit data.  Then, the new symmetrized data point is set equal to
the weighted mean of the points on both sides, with weights
proportional to the inverse square of the original errors.  If
$\sigma_+$ and $\sigma_-$ are the errors on both sides, the new error
$\sigma$ on the symmetrized data points is set equal to the maximum of
\begin{equation}
\frac{2}{\sigma^2} = \frac{1}{\sigma^2_+} + \frac{1}{\sigma^2_-}
\end{equation}
and half of the deviation between the original data points,
which includes systematic errors between both sides of the galaxy.

The resulting data, with their respective error bars, are shown with
red dots in Fig.~\ref{fig:slit_data}, where they can be compared with
the original measurements.

\subsubsection{Stellar-absorption line slitlets data}
\label{sec:lets}
\begin{figure*}
\includegraphics[angle=90.0,scale=0.75]{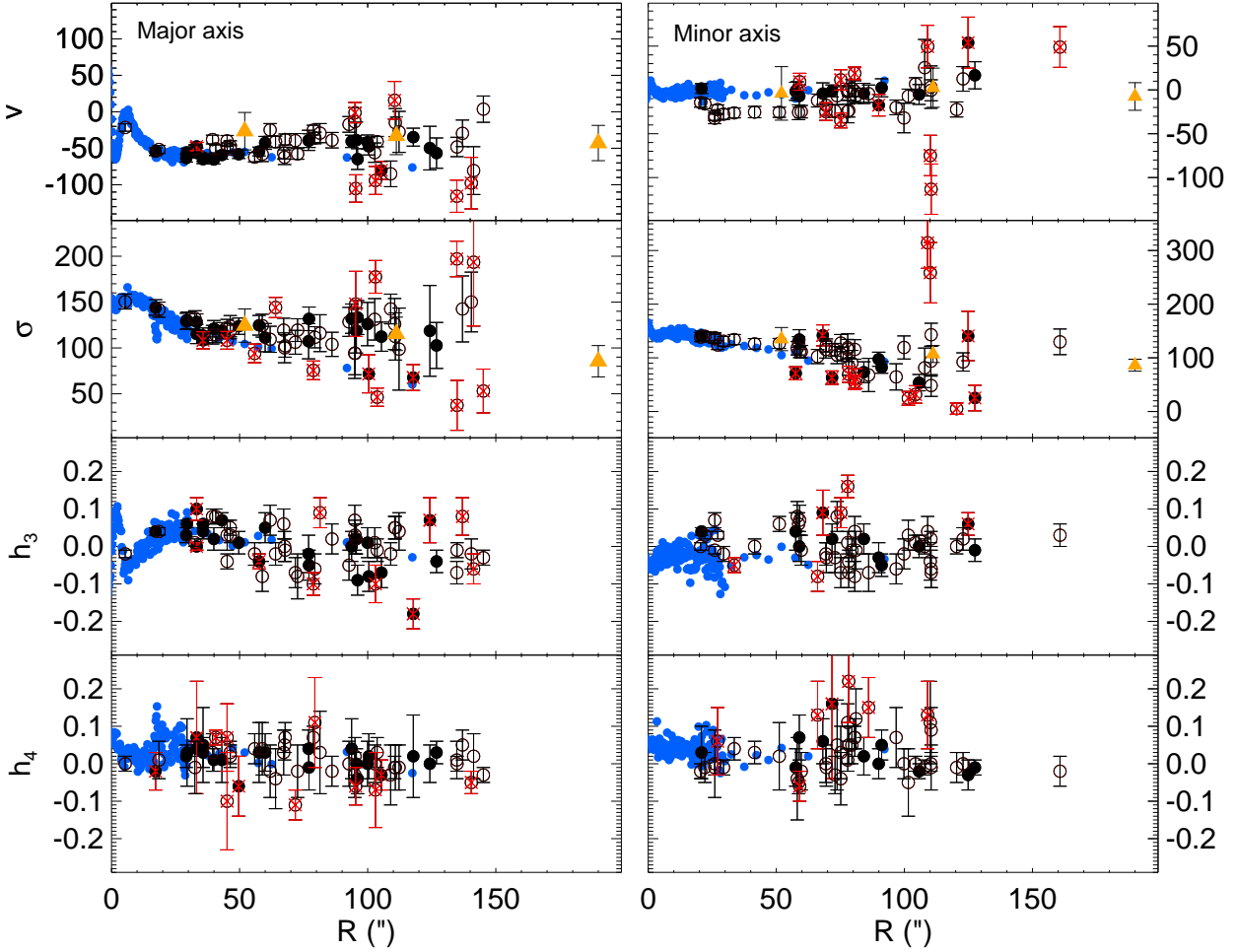}
\caption[]{Slitlets kinematics in the 4-folded data set ({\it
    circles}), compared with long-slit ({\it blue dots}) and PNe data
  ({\it orange triangles}).  Left and right column are for the
  measurements around the major and minor axis, respectively.  {\it
    Filled circles} represent the data in a cone of $\pm15\degrees$
  around the axis, while {\it empty circles} are the data points in
  cones $30\degrees\pm15\degrees$, $60\degrees\pm15\degrees$,
  respectively.  Outliers determined by the method described in
  Section~\ref{sec:lets} are marked {\it red}.  }
\label{fig:lets}
\end{figure*}

We include in our kinematic constraints the measurements of $v$,
$\sigma$, $h_3$, and $h_4$ in 115 galaxy light spectra in slitlets
recently published by \citetalias{foster11}, which extend out to
$3.5R_{\rm e}$.

\citetalias{foster11} discussed the generally good agreement between
these data and the long-slit absorption line kinematics of
\citet{coccato09}.  Also, they reported that their uncertainties are
likely to be slightly underestimated.

On the whole, the absorption line kinematic data do not show
significant evidence for minor axis rotation.  Therefore, being
interested in axisymmetric models of the data, we 4-fold the original
sample of slitlets in order to decrease the impact of data asymmetries
in our models. To this end we reflect the measurement points
  around both the galaxy's major and minor axes, maintaining errors,
  and taking into account the sign reversal of $v$ and $h_3$.

Finally, we look for possible outliers in the 4-folded data set
consisting of 460 data points.  To this aim, we use the following
procedure: for each slitlet, we compute the value of the average field
of $v$, $\sigma$, $h_3$, and $h_4$ from its 20 nearest neighbours,
excluding the slitlet itself.  This average is worked out by excluding
the lowest and highest 2 values in the neighbourhood, i.e. considering
only the central $80\%$ of the distribution.  Moreover, the average is
a weighted one, with weights equal to the inverse square of the larger
of the individual observational error and the median of the
observational errors in the central $80\%$ of the distribution.  In
the same way, we calculate a weighted rms $\sigma$ in each
neighbourhood.  Then, we flag as outlier any point deviating by more
than $2\sigma$ from the weighted mean of its neighbours.

With this procedure, 158 data points are flagged as outliers in
velocity and/or velocity dispersion, and removed from the 4-folded
data set.  Of the resulting data set consisting of 302 slitlets, 68
points are marked as outliers in $h_3$ or $h_4$, and therefore we only
consider their velocity and velocity dispersion in the modelling
below.

The slitlets data points, together with the outliers determined in
this way, are shown in Fig.~\ref{fig:lets}.

\subsubsection{Planetary nebulae velocities}
\label{sec:pne}

Our kinematic observables are completed by the 267 PNe line-of-sight
velocities obtained by \citetalias{napolitano09} with the Planetary
Nebulae Spectrograph \citep[PNS,][]{pns}.  As illustrated in
Fig.~\ref{fig:setup}, PNe extend out to $\sim7R_{\rm e}$.

Fig.~\ref{fig:lets} shows the comparison of all kinematic data.  In
the region of overlap, the kinematics of PNe are consistent with those
of the stars \citepalias[see][Fig.~6 therein]{napolitano09}.

\begin{figure}
\centering
\includegraphics[height=7cm,angle=-90.0]{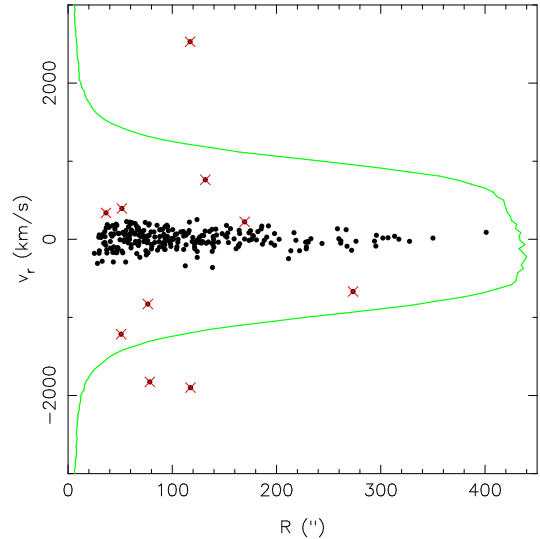}
\caption[]{Distribution of the line-of-sight velocities of PNe as a
  function of projected radius. {\it Red crosses} mark the outliers
  identified by the friendless algorithm. The PNS filter band-pass is
  overplotted in {\it green}.}
\label{fig:outliers}
\end{figure}

In order to look for possible outliers in the sample of PNe, we use
the friendless algorithm presented in \citet{mer03}, which flags as
outlier any object deviating by more than $n\times\sigma$ from the
velocity distribution of its $M$ nearest neighbours, where $\sigma$ is
the rms computed in each neighbourhood.  Adopting $n=2.5$ and $M=20$,
we remove 10 outliers from the original sample of PNe.  These outliers
are highlighted in Fig.~\ref{fig:outliers}, which shows the projected
phase-space distribution of PNe.

We are left with a catalogue of 257 PNe, whose size we double by
applying point-symmetry, i.e. generating for every PN (x,y,v) the
symmetric PN (-x,-y,-v).  Such point-symmetric velocity fields are
expected in axisymmetric (or triaxial) potentials.  From previous
experiments we know that increasing the number of PNe by a further
factor of two does not improve the modelling significantly.

\section{MODELLING DATA WITH NMAGIC}
\label{sec:mod}
In this Section we briefly describe the NMAGIC made-to-measure (M2M)
modelling technique.  We also explain how the initial particle model
is set up, and how the photometric and kinematic target observables
are preprocessed for the dynamical modelling.

\subsection{NMAGIC}
\label{sub:nmagic}
The parallel code NMAGIC \citepalias{dl07} is an implementation of a
particle-based method to create $\chi^2$M2M models in agreement with
observations of galaxies.  The algorithm is a slight modification of
the technique proposed by \citet{st96}, designed to model
observational data with errors.

The basic idea behind M2M particle methods is to train a system of
$i=1,\dots, N$ particles to reproduce the observables of a target
galaxy by maximizing the function
\begin{equation}
F = -\frac{1}{2}\chi^2+\mu S+\mathcal L
\label{eqn:SF}
\end{equation}
with respect to the particle weights $w_i$.  This maximization strikes
a compromise between the goodness of the fit $(\chi^2)$ in terms of
deviations between target and particle model observables, and a
pseudo-entropy functional
\begin{equation} 
S=-\sum_{i=1}^N w_i\left[\log\left(\frac{w_i}{\tilde{w_i}}\right)-1\right]
\label{eqn:entropy}
\end{equation}
\citepalias{mg11} which serves the purpose of regularization.
Finally, the likelihood term $\mathcal L$ is added to
equation~(\ref{eqn:SF}) to account for the likelihood of a sample of
PNe velocities \citepalias[see][]{dl08}.

Since in typical applications the number of particles is much higher
than the number of data constraints on the particle model,
regularization is essential. In standard M2M, regularization is
achieved by pushing the individual particle weights towards a smooth
distribution of predetermined priors $\tilde{w_i}$, which mirror the
initial particle weight distribution and are kept constant during the
modelling.  As we showed in \citetalias{mg11}, such a Global Weight
Entropy scheme makes it hard to reconcile smoothness and orbital
anisotropies in the final particle model.  Therefore, in this work we
adopt the alternative Moving Prior Regularization (MPR) method
proposed in \citetalias{mg11}, and we determine priors which follow
the smooth phase-space structures traced by the weight distribution,
as the latter adapts to match the observational data.  This method
facilitates recovering both a smoother and more accurate mass
distribution function from noisy data, smoothing over local
fluctuations without erasing global phase-space gradients.

Maximizing the function~(\ref{eqn:SF}) translates into a prescription,
the so-called ``force-of-change'' equation, for correcting the weights
of the particles while these are evolved in the total gravitational
potential. In generic form this is
\begin{eqnarray}
\frac{\ud w_i(t)}{\ud t} = - \varepsilon w_i(t)
  \Bigg[&&\sum_{\rm id} \xi_{\rm id} \sum_{j=1}^{J_{\rm id}}  
                 \frac{K_j^{\rm id} \left[{\mathbf z}_i(t)\right]} {\sigma_j^{\rm id}}
             \;\widetilde{\uDelta}_j^{\;\rm id}(t) \nonumber \\
        && -\mu \frac{\partial S}{\partial w_i} 
        -  \xi_{\rm PN} \frac{\partial}{\partial w_i} \sum_{j=1}^{J_{\rm PN}} \ln\mathcal L_j
  \Bigg].
\label{eqn:myFOC}
\end{eqnarray} 
The first term on the right is the $\chi^2$ derivative term as in
\citetalias{dl07}, summed over the observational constraints for the
different types of data observables, ${\rm id}$: $A_{\rm lm}$, slits,
and slitlets; $J_{\rm id}$ are the respective number of constraints.
$\widetilde{\uDelta}_j^{\;\rm id}$ and ${\sigma_j^{\rm id}}$ are the
temporally-smoothed relative deviation and error in observable $({\rm
  id},j)$, and $K_j^{\rm id}$ the corresponding kernel evaluated at
phase-space position ${\mathbf z}_i(t)$ of particle $i$. The last term
has been written as sum over the log likelihoods as in
\citetalias{dl08}. The notation is explained further in these papers.
We have included here additional separate weight factors $\xi_{\rm
  id}$ for the different observables, which we use to optimize the
influence of the data on the evolution of the model \citep[see
also][]{longmao10}.  

\subsection{The gravitational potential}
\label{sub:vc}
The particle modelling technique allows for using both a fixed
potential, known a priori, and a time-varying potential,
self-consistently computed from the particle distribution.  In our
dynamical models, we assume that the total gravitational potential is
generated by the luminous and dark matter distributions, i.e.
\begin{equation}
\phi = \phi_{\star}+\phi_D.
\label{eq:phi}
\end{equation}

Following \citet{sellwood03} and \citetalias{dl07}, $\phi_{\star}$ is
frequently computed from the $N$-particle model for the light
distribution via a spherical harmonic decomposition, assuming a
constant mass-to-light ratio $\Upsilon$.  The value of $\Upsilon$ is
not a fixed parameter, but rather it is determined during the NMAGIC
run, simultaneously with the modelling of the observational data
\citepalias[see][]{dl08}.

Instead, the dark matter halo potential $\phi_D$ is parametrized,
and has the logarithmic form
\begin{equation} 
 \phi_D(R,z)=\frac{v_0^2}{2}\ln\left(r_0^2+R^2\right)
\label{eq:log_halo}
\end{equation}
\citep{bt08}, where $v_0$ and $r_0$ are a characteristic (constant)
circular velocity and scale-length.  This mass model has been widely
and successfully used to fit galaxies
\citep[\eg][\citetalias{dl09}]{fall80,persic96,kron00,thomas07}.

\begin{figure}
\includegraphics[scale=0.5,angle=0.0]{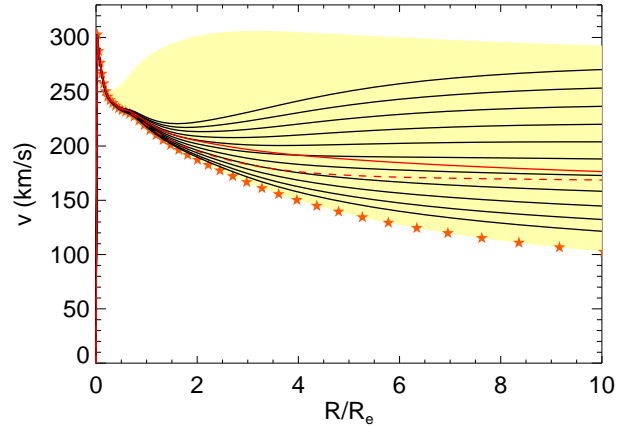}
\caption[]{The {\it shaded area} shows the range of circular velocity 
curves corresponding to the total gravitational potentials used in the 
dynamical models. 
{\it Stars} represent the self-consistent case with constant $M/L$;
{\it solid lines} correspond to models embedded in various spherical dark 
matter halos. 
{\it Black} lines correspond to dark halos with $r_0=4R_{\rm e}$ and 
$v_0=[70,...,270]\;{\rm km s}^{-1}$ (from bottom to top).
{\it Solid} and {\it dashed red} lines correspond to 
$v_0=150{\rm km s}^{-1}$, and $r_0=3R_{\rm e}$ and $5R_{\rm e}$, respectively.}
\label{fig:vc}
\end{figure}

Our dynamical models will explore a range of circular velocity curves,
whose behaviour at large radii varies between the near-Keplerian
decline (when stars dominate the total potential), and a nearly flat
(quasi-isothermal) shape obtained for massive dark halos, as shown in
Fig.~\ref{fig:vc}.

\subsection{The initial particle model}
\label{ssec:ics}
We set up initial models of $N=750000$ particles extending 
to $30R_{\rm e}$, i.e. $\sim110$ kpc at the distance of NGC~4494.

The density of these initial particle models is given by a spherical
deprojection of the circularly-averaged surface brightness profile.
We follow the method described in \citet{gerhard91} to obtain an
isotropic stellar velocity distribution in the total gravitational
potential generated by the stars plus the different dark matter halos
described above.

The particles' coordinates and velocities are drawn according to the
complete distribution function applying the method of
\citet{debatsel00}, and particle weights are set equal to $1/N$.

Finally, following \citet{kal77} and \citetalias{dl08}, we switch the
sign of the velocity of a fraction of the retrograde particles, with
probability
\begin{equation}
 p(L_z)=p_0\frac{L_z^2}{L_z^2+L_0^2}
\end{equation}
(where $p_0=0.3$, $L_0=0.02$), to introduce some angular momentum
about the $z$-axis, while maintaining a smooth DF in equilibrium.
This expedient makes it easier to reproduce the rotation velocity seen
in the kinematic data.

\subsection{Photometric and kinematic observables for the modelling}
\label{sec:obs_nmagic}
We now explain how the different photometric and kinematic
observational data are processed in order to be used as constraints
for the NMAGIC models.

For the photometric observables, the 3D luminosity density profile
obtained from the deprojection of the surface-brightness is expanded
in spherical harmonic functions, and the expansion coefficients
$A_{lm}$ are used as luminosity constraints \citepalias[see][]{dl07}.
These $A_{lm}$ are computed on a grid of 50 quasi-logarithmically
spaced radial shells between $r_{\rm min}=0.01''$ and $r_{\rm
  max}=20R_{\rm e}$, and all terms that would allow non-axisymmetry
are set to zero.  Poissonian errors are used for the mass in shells,
whereas the errors for the higher-order mass moments are determined
via Monte Carlo simulations of the density field of the target galaxy
\citepalias[see][]{dl07}.

For the long-slits and slitlets data, we use luminosity-weighted
kinematic observables.  Hence, we add to the target observables the
luminosity in each slit cell or slitlet, obtained by integrating the
surface brightness distribution with a Monte Carlo algorithm.  For the
slits, we assume that the slitwidth equals $5''$; for the slitlets, we
set the slitwidth equal to the larger value between $1''$ and the
diameter of a circle containing at least $250$ particles of the
initial particle model, so as to limit particle noise when computing
model observables. These slit cell luminosities are independently
fitted, with errors set to $1\%$ of the luminosity in each cell.

\begin{figure}
\centering
\includegraphics[height=6cm,angle=-90.0]{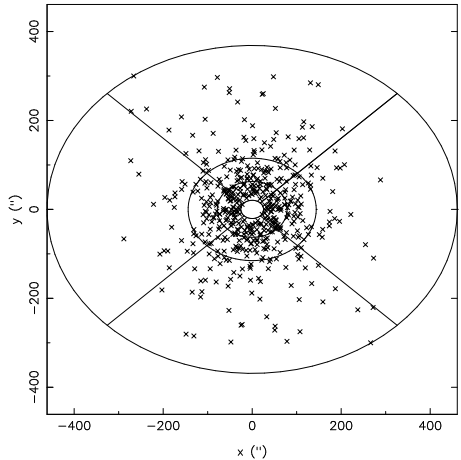}
\caption[]{Segments in which the line-of-sight velocity distribution
  of particles is computed, for the likelihood method used to fit PNe
  with NMAGIC.}
\label{fig:pne_bins}
\end{figure}

PNe data are modelled by maximizing the likelihood of the sample of
discrete velocities and positions, as detailed in \citetalias{dl08}.
For computing the likelihood, the particles and PNe are binned in
elliptical segments, assuming an average projected ellipticity of 0.2.
We consider 3 radial and 4 equally-spaced angular bins, with the first
angular bin centered on the major axis, as shown in
Fig.~\ref{fig:pne_bins}.  Each segment contains at least $30$ PNe.

The number of independent data constraints for the different
  observables are $J_{\rm A_{lm}}=450$, $J_{\rm Sl}=2390$, $J_{\rm
    Lets}=346$, and $J_{\rm PN}=257$.  Since the errors are smallest
  for the $A_{lm}$, while the Poisson errors for the PNe are large,
  the $A_{lm}$ terms contribute most to the force-of-change equation,
  while changes to the particle weights due the PNe data are small.
  To ensure that the halo is appropriately modelled, we increase the
  weight of the slitlets in the force-of-change by a factor of 4, and
  that of the PNe by a factor of 20.  In practise we use the 4-folded
  data set for the slitlets (\S\ref{sec:lets}) and the 2-folded data
  set for the PNe (\S\ref{sec:pne}), so the actual weight factors in
  equation~(\ref{eqn:myFOC}) are $\xi_{\rm A_{lm}}=\xi_{\rm Sl} = 
  \xi_{\rm Lets}=1$, and $\xi_{\rm PN}=10$.  

\subsection{Modelling procedure}\label{sec:modelling}
Starting from the initial particle model described in
Section~\ref{ssec:ics}, the weights of all particles are evolved by
NMAGIC until the $N$-body system matches the target.  During the whole
evolution, the potential of the dark halo is kept fixed, but the
stellar potential is frequently computed from the light distribution
of the particle system.  Particles are integrated in the total
gravitational potential using a leapfrog scheme with adaptive time
step.

After a relaxation phase of 1000 steps in which the initial particle
model is advanced without any weight correction, particle weights are
updated according to the force-of-change obtained maximizing
equation~(\ref{eqn:SF}), for $\sim10^5$ correction time steps.  Finally,
the particles are freely evolved for another $10^4$ steps without any
further weight correction, to ensure that the final particle model is
well phase-mixed.  For reference, $10^4$ correction time steps
correspond to $\sim300$ circular rotation periods at the target
$R_{\rm e}$.  We define the model to have converged if $\chi^2$
averaged over 50 steps stays almost constant in the last $10^4$ steps,
with fluctuations which are typically of order 2\%.

During the weight adaptation, the model is regularized by the
MPR method proposed in \citetalias{mg11}. New priors are
determined in phase-space as described in \citetalias{mg11}, by
binning particles according to their orbital integrals.  In
particular, we grid particles according to energy ($E$), total
circularity $x$, and angular momentum with respect to the
rotation axis $L_z$, in a grid of $n_E=20$, $n_x=4$, $n_z=2$ bins,
which at the same time resolves the relevant phase-space structures
and ensures enough particles in each grid cell.  The average weight
contained in each cell is computed, and then a thin-plate smoothing
spline is fitted to the grid of priors, before assigning them to the
particles.  The individual particle priors computed in this way
are not kept constant in time but rather they are frequently updated
while the particle weights are adapted to match the target
observables.

\section{Parameter estimation with NMAGIC: how well can the dark
  matter halo be recovered?}
\label{sec:test}

In this work, we will be fitting the observational data of NGC 4494
with a sequence of NMAGIC models obtained for different dark matter
halos and for different inclinations.  In doing this, we wish to
quantify the uncertainties in the best-fit parameters from a
statistical distribution appropriate for our data and modelling
method. In this Section, we describe a simple Monte Carlo method to
estimate such a distribution.

In the following, the best-fit parameters are defined as those
for which a minimum is achieved in the merit function
\begin{equation}
 G=\frac{1}{2}\chi^2-\mathcal L,
\label{eq:G}
\end{equation}
which measures the agreement between the  independent
observational data constraints and the final model in terms of
the total $\chi^2$ from $A_{lm}$, slits, and slitlets
that the $\chi^2$M2M method tries to minimize, and of the log
likelihood $\mathcal L$ for the PNe that the method tries to
maximize.  In order to carry out a robust parameter estimation, we
also need to quantify the uncertainties in these best-fit parameters.

Boundaries of the confidence regions for the estimated parameters will
correspond to contours of constant $\Delta G$ relative to the
best-fitting model in the sampled parameter space.  In common-practice
dynamical mass analysis (see Introduction), uncertainties in the model
parameters are typically estimated by considering the parameter change
necessary to increase $\chi^2$ by 1 (or 2.3), and to decrease the log
likelihood $\mathcal L$ \citepalias[see][]{dl08} by a factor of 0.5
(or 1.2) for 1 (or 2) fitted potential parameters \citep{press92}.
This assumes Gaussian data, a model linear in its parameters, and
  that the number of fitted weight parameters $N$ minus smoothing
  constraints $N_s$ is less than the number of data points $J$ so that
  the number of degrees-of-freedom $N_{\rm dof}$ is positive.  In
  fact, the dependence on the potential parameters is non-linear, and
  in our models $N\gg J$ so that $N_{\rm dof}$ may be less than zero.
  In addition, we find empirically for our data set and modelling
set-up that $\Delta\chi^2$ values are $\gg1$. Therefore in this
Section we compute confidence regions for parameter estimation via
Monte Carlo experiments.

In Section 4.1, we build a mock target galaxy closely similar to the
elliptical galaxy NGC 4494, which is embedded in a dark matter halo
with scale radius and circular velocity corresponding to the
best-fitting model determined by N09 for NGC 4494. We generate
photometric and kinematic observables for this mock galaxy with the
same observational errors as for NGC 4494.  In Section 4.2, we use the
pseudo-data and known intrinsic parameters of this NGC 4494-like
target galaxy to calibrate the regularization parameter appropriate
for the modelling of NGC 4494.  In Section 4.3, we determine how well
the circular velocity and scale radius of the known dark halo, the
total enclosed mass, and the mass-to-light ratio of the target galaxy
can be recovered, in a more rigorous way than in previous NMAGIC
modelling \citep[\citetalias{dl08,dl09};][]{payel2}.  Finally, in
Section 4.4 we use Monte Carlo experiments with such NGC~4494-like
target galaxies to determine the appropriate $\Delta G$ to be used for
confidence boundaries. This will then allow us to quantify
uncertainties in the dark halo mass distributions of the real NGC~4494
in Section~\ref{sec:models}.

\subsection{An NGC~4494-like galaxy and its observables}
\label{ssec:pseudo}
Our initial NGC~4494-like target galaxy has the luminosity distribution
obtained by deprojecting the surface brightness of NGC~4494 for
$i=90\degrees$, and it is embedded in a logarithmic dark matter halo
(equation~(\ref{eq:log_halo}), with $r_0/R_e=4$ and $v_0=150{\rm km s}^{-1}$), 
as in the best-fitting Jeans model of \citetalias{napolitano09} for
NGC~4494.  Similar to the orbital anisotropy predicted by some merger
models in the current cosmological scenario \citep[\eg][]{dekel05},
the velocity distribution of this NGC~4494-like galaxy is isotropic in
the center and increasingly radially anisotropic at larger radii.  As
for NGC~4494, we observe the target galaxy from a distance of
$15.8$~Mpc, and the projected effective radius $R_{\rm e}\approx
49''$.  We set its stellar mass-to-light ratio to $3.8$.

Following the procedure outlined in Section~\ref{ssec:ics}, we
generate a spherical particle model realization for this target with 750000
particles.  To implement the orbital anisotropy of the target galaxy,
we adopt a model with specified circularity function following
\citet{gerhard91}.  We also add a certain amount of rotation to the
particle model (see Section~\ref{ssec:ics}) so as to mimic the real
NGC~4494. With this spherical model as starting point, we then
use NMAGIC to evolve it to the edge-on luminosity distribution of
NGC~4494. The resulting axisymmetric model galaxy is then used to
compute mock observables.

As luminosity observables we use the coefficients $A_{lm}$ of the
spherical harmonic expansion of the 3D luminosity density, computed as
described in Section~\ref{sec:obs_nmagic}.  The kinematic observables
are $v,\sigma,h_3,h_4$, projected onto both the long-slit set-up and
the slitlets set-up used for the modelling of NGC~4494.  These
observables can be readily determined using NMAGIC with weight
correction turned off, integrating the particle model for the
target galaxy, and time-smoothing the observables
\citepalias[see][]{mg11}.  Observational errors from NGC~4494 are
adopted, and Gaussian random variates with $1\sigma$ equal to these
errors are added to the kinematic observables computed in this way.

Finally, we generate a mock sample of PNe velocities similar to that
described in Section~\ref{sec:pne}.  We again use NMAGIC to integrate
the particle model which imitates NGC~4494, and in parallel compute
the time-averaged velocity and velocity dispersion of the particles
binned in radial and angular segments on the sky.  We adopt the same 3
radial bins and 4 equally-spaced angular bins as in
Section~\ref{sec:obs_nmagic}.  Then, we consider the catalogue of PNe
used for the modelling of NGC~4494, and assign a new velocity to every
PN in the catalogue, according to a Gaussian distribution with
velocity and velocity dispersion of the spatial segment the PN would
belong to, given its position on the sky.

%%%%%%%%%%%%%%%%%%%%%%%%%%%%%%%%%%%%%%%%%%%%%%%%%%%%%%%%%%%%%%%%%%%%%%%%%%%%%%%%%%%%%%%%%%%
\begin{figure}
\centering
\includegraphics[height=6cm,angle=-90.0]{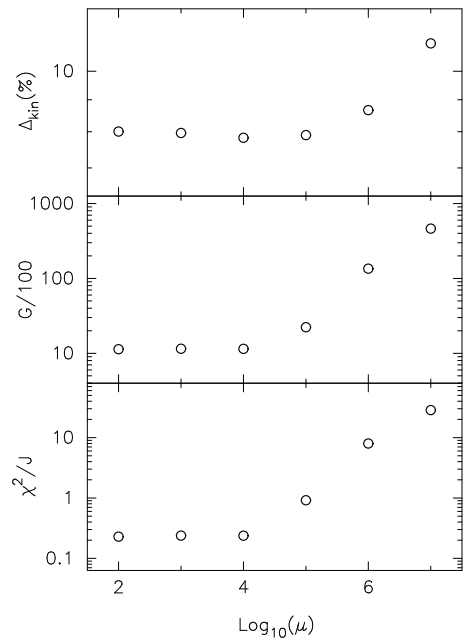}
\caption[]{Quality parameters for the final NMAGIC particle models as
  a function of the regularization parameter $\mu$, from unsmoothed
  models (small values of $\mu$) to oversmoothed models (high values
  of $\mu$).
\textit{Top panel}: rms deviation (\%) of the internal velocity moments
of the final particle model from the target, computed over the radial 
region constrained by the data.
\textit{Middle panel}: merit function $G$ in equation~(\ref{eq:G}).
\textit{Bottom panel}: $\chi^2/J$ of the particle model fit to 
the $J=3186$ photometric and absorption-line kinematic data points.
}
\label{fig:mu}
\end{figure}
%%%%%%%%%%%%%%%%%%%%%%%%%%%%%%%%%%%%%%%%%%%%%%%%%%%%%%%%%%%%%%%%%%%%%%%%%%%%%%%%%%%%%%%%%%%

\subsection{Calibrating regularization}
\label{ssec:mu}
The regularization parameter to be used when constructing dynamical
models is case-dependent, and is influenced by several factors, most
notably the observational data to be modelled (error bars, scatter,
spatial coverage) and the phase-space structure of the target galaxy
\citep[see][\citetalias{dl08,dl09,mg11}]{gerhard98,cretton99,thomas05}.

To determine the optimal amount of smoothing given the data, we model
the observational data of the NGC~4494-like galaxy using MPR
(\citetalias{mg11}), varying the regularization parameter $\mu$ which
controls the balance between regularization and goodness-of-fit in
equation~(\ref{eqn:SF}).

The results of our experiments are summarized in Fig.~\ref{fig:mu},
which shows, for increasing values of $\mu$, the reduced $\chi^2$, the
merit function $G$ in equation~(\ref{eq:G}), and the rms difference
($\Delta_{kin}$) between the internal velocity moments of the target
and of the final NMAGIC model \citepalias[see][]{mg11}.

The minimum in the $\Delta_{\rm kin}$ plot determines the value of
$\mu$ for which the model best recovers the internal moments of the
input model.  This occurs at $\mu\simeq 10^{4}$.  In Fig.~\ref{fig:mu}
we also see that for values of $\mu\geq10^5$ a smooth model fitting
the data points within errors can no longer be found.

Based on the results of these experiments, in the following we will
adopt a value of $\mu=10^{4}$ to regularize our NMAGIC models.

\subsection{Recovery of input parameters for the dark halo of an
  NGC~4494-like galaxy}

We now construct NMAGIC models that fit the data of the NGC~4494-like
target galaxy for a range of assumed parameters of the dark halo
potential in equation~(\ref{eq:log_halo}).  The key question that we
are interested in is: how well can NMAGIC recover the true parameters
of the dark halo given the observational data at hand?

\begin{figure}
\centering
\includegraphics[scale=0.77,trim=30 20 10 20,angle=0.0]{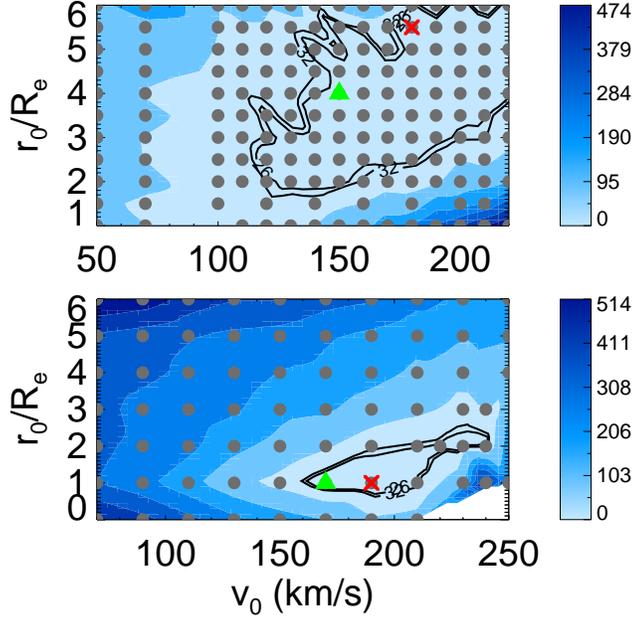}
\caption[]{Recovering the halo parameters of NGC~4494-like
    galaxies.  Each {\it circle} represents a final NMAGIC model.
    {\it Top panel:} the true dark halo ({\it green triangle}) has
    scale radius $r_0=4R_{\rm e}$ and circular velocity $v_0=150{\rm
      km s}^{-1}$.  {\it Bottom panel:} this more massive dark halo
    ({\it green triangle}) has scale radius $r_0=1R_{\rm e}$ and
    circular velocity $v_0=170{\rm km s}^{-1}$.  The colour scheme
    reflects the magnitude of the $\Delta G$ of each model relative to
    the overall best-fitting model, defined as that with the smallest
    $G$ ({\it red cross}).  The {\it black lines} correspond to 70\%
    and 90\% confidence contours, as derived from the simulations of
    Section~\ref{ssec:mc}.}  
\label{fig:r4v150_delg}
\end{figure}

We follow the modelling procedure outlined in
Section~\ref{sec:modelling}, starting from an isotropic initial
particle model computed in each dark halo, as explained in
Section~\ref{ssec:ics}.  We use all available photometric and
kinematic constraints, including PNe through the likelihood technique.
Our NMAGIC models are regularized with MPR, adopting the optimal value
of $\mu=10^4$ determined above.

We sample the halo parameters ($r_0$,$v_0$) on a grid of $\sim200$
models.  The results of these experiments are shown in the top
  panel of Fig.~\ref{fig:r4v150_delg} in terms of the $\Delta G$ of
each NMAGIC model relative to the best-fitting model, which
corresponds to the overall smallest $G$.  The shape of the $\Delta G$
contours is regular, and there is an extended region for which the
values of $G$ are similar. Many models for different
  parameters of the dark halo provide similarly good fits to the data,
  all with $\chi^2/J\leq1$ (the number of data points entering
  $\chi^2$ is $J=3186$).

The best-fitting model is found for values of the dark halo
  parameters $r_0=5.5R_{\rm e}$ and $v_0=180{\rm km s}^{-1}$, and it
  has $\chi^2/J=0.22$, log likelihood $\mathcal L=-770.5$, and
  $G=1125.6$.  The scale radius of this best-fitting model is larger
  than that of the true model, which achieves $\chi^2/J=0.24$, log
  likelihood $\mathcal L=-769.9$, and $G=1148.7$.  For comparison, the
  no-dark halo model, which provides the poorest fit to the available
  observational data, has a final $\chi^2/J=0.37$, log likelihood
  $\mathcal L=-782.2$, and $G=1376.6$.

Surprisingly, the values of $\Delta G$ are large $(\gg1)$.
For instance, the model obtained for the true parameters of
  the (known) dark matter halo achieves a value of $G$ which is
  $\Delta G\sim20$ ($\Delta \chi^2\sim 60$) away from the best-fitting
  model. Applying the classical $\chi^2$ statistics for two degrees
of freedom, the true parameters of the dark matter halo would be
discarded at the 99.99\% level.  We investigate this issue in
Section~\ref{ssec:mc}.

Note that the contours of $\Delta G$ remain open to the top-right edge
of the plot \citep[see also \eg][]{gerhard98,thomas05,murphy11},
showing that it is not possible to put robust constraints on both halo
parameters simultaneously in this case.  Similar values of $G$, and
hence similarly good models are achieved for halo models located along
a large band extending from low $r_0$ and low $v_0$ to high $r_0$ and
high $v_0$.

This diagonal band shrinks when modeling target galaxies embedded in
more massive halos, as the models that do not contain enough mass are
then ruled out. This is shown in the bottom panel of
  Fig.~\ref{fig:r4v150_delg}, which is obtained by repeating the
  analysis for an NGC~4494-like galaxy embedded in a more massive dark
  matter halo (whose parameters are set to $r_0=1R_{\rm e}$ and
  $v_0=170{\rm km s}^{-1}$).

\begin{figure}
\includegraphics[scale=0.5,angle=0.0]{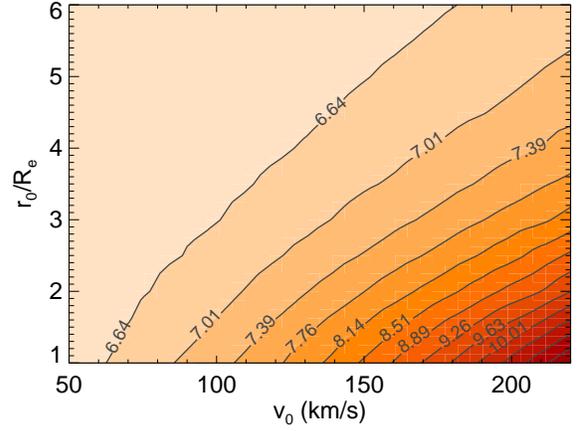}
\caption[]{Total enclosed mass within $2R_{\rm e}$ (stars plus dark
  matter) for different values of the halo parameters $r_0$ and $v_0$.
  Contour values are in units of $10^{10}M_{\sun}.$}
\label{fig:mass}
\end{figure}

\begin{figure}
\includegraphics[scale=0.5,angle=0.0]{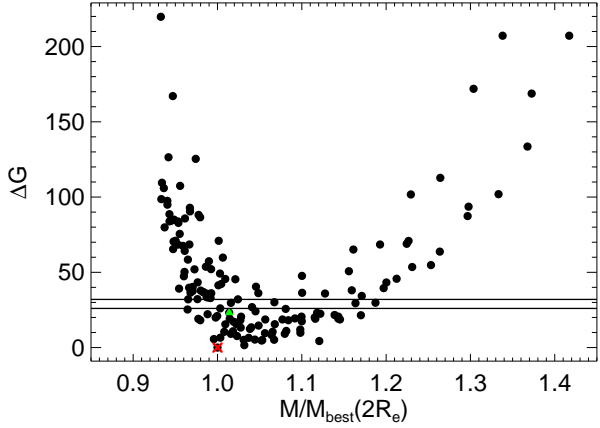}
\caption[]{$\Delta G$ of the final NMAGIC models 
fitting the data of the NGC~4494-like galaxy, as a function of
the total mass inside $2R_{\rm e}$ 
divided by the corresponding quantity for the best-fitting model.
The {\it red cross} shows the best-fitting model,
the {\it green triangle} the model with the true dark halo parameters
($r_0=4R_{\rm e}$, $v_0=150{\rm km s}^{-1}$).
The black lines correspond to 70\% and 90\% confidence regions
derived from the simulations of Section~\ref{ssec:mc}.
}
\label{fig:r4v150_mass}
\end{figure}

Models inside the diagonal band share a similar total enclosed mass
within the radial region constrained by the observational data, as
shown in Fig.~\ref{fig:mass} by means of contours of total mass within
$2R_{\rm e}$ in the same ($r_0$,$v_0$) parameter space.  Evidently,
the total enclosed mass inside a certain radius is what the dynamical
models constrain best with the data at hand.  Indeed, when plotting
the values of $\Delta G$ of the final NMAGIC models against enclosed
mass (see Fig.~\ref{fig:r4v150_mass}) the shape of a parabola is
evident, despite the scatter in the values of $\Delta G$.

\begin{figure}
\includegraphics[scale=0.5,angle=0.0]{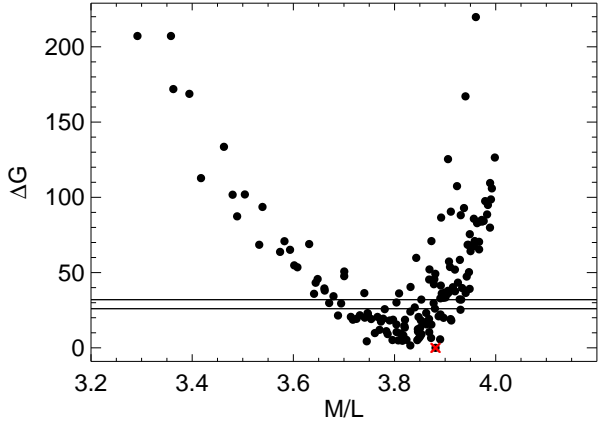}
\caption[]{As Fig.~\ref{fig:r4v150_mass}, but $\Delta G$ of the final
  NMAGIC models as a function of stellar mass-to-light. The galaxy has
  (known) $M/L=3.8$.  }
\label{fig:m_to_l_mock}
\end{figure}

\begin{figure}
\includegraphics[scale=0.5,angle=0.0]{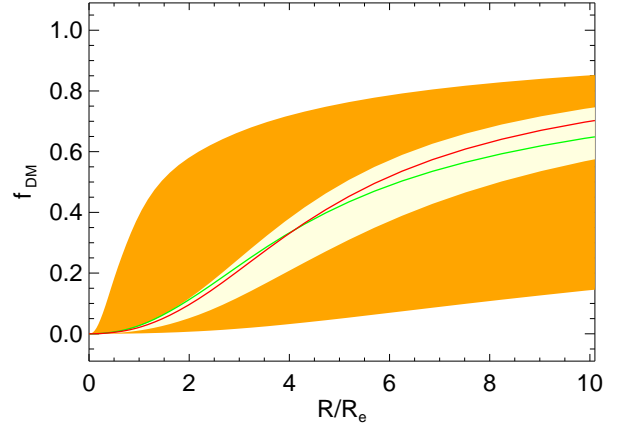}
\caption[]{Recovery of the dark matter fraction
$f_{\rm DM}=M_{\rm{DM}}/M$ as a function of radius 
for the range of explored NMAGIC models 
of the NGC~4494-like data ({\it orange shaded region}).
The {\it green line} represents the known target,
and the {\it red line} the best-fitting model.
The {\it yellow shaded region} shows the 70\% confidence region
(see Section~\ref{ssec:mc}).
}
\label{fig:fdm_mock}
\end{figure}

Figs.~\ref{fig:m_to_l_mock}-\ref{fig:fdm_mock} show how well the
NMAGIC models recover the mass-to-light ratio and dark matter fraction
of the NGC~4494-like galaxy from the mock data.  These are both
important quantities which we are interested in measuring with
accuracy for real galaxies.

At this stage, two obvious questions are: what is the typical $\Delta
G$ difference between the best-fitting NMAGIC model and the model with
the true halo parameters? And related to this, within what errors can
we trust the dark halo parameters of the best-fitting model? The
following analysis is designed to answer these questions.

\subsection{Parameter estimation for the dark matter halo: confidence levels}
\label{ssec:mc}
\begin{figure}
\centering
\includegraphics[height=5cm,angle=0.0]{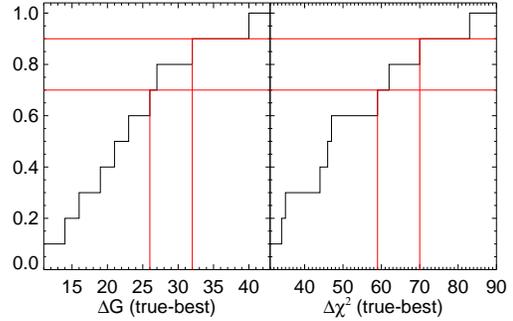}
\caption[]{Normalized cumulative distribution of $\Delta G$ ({\it
    left}) and $\Delta\chi^2$ ({\it right}) between the true dark halo
  and the best fitting model, derived from NMAGIC models of the
  NGC~4494-like galaxy embedded in 10 different dark halos.  The best
  fitting model is defined as that with the minimum $G$ ($\chi^2$).
  The {\it red lines} correspond to 70\% and 90\% confidence levels. }
\label{fig:cdf}
\end{figure}  
     
We use Monte Carlo simulations to estimate the values of $\Delta G$
which represent a specific confidence level, given the observational
data of NGC~4494 and our NMAGIC modelling technique.

In practice, given constraints on computer time, we construct a
sequence of NGC~4494-like galaxies embedded in 10 different dark
halos, and model each one of them with a range of dark halo
parameters.  Then, for each target galaxy we compute the $\Delta G$
between the best-fitting model and the model obtained for the true
parameters.  In this way, we can estimate the values of $\Delta G$
within which the true parameters are found in 70\% or 90\% of our
experiments.

The main results from these experiments are summarized in
Fig.~\ref{fig:cdf}, which shows the cumulative distributions of
$\Delta G$ and $\Delta\chi^2$ between the true dark halo model and the
best fitting model, for the 10 different experiments. It is
  seen that the NMAGIC models recover the true potential of the dark
  matter halo within $\Delta G\lt26$ about 70\% of the time, and
  within $\Delta G\lt32$ about 90\% of the time.  Corresponding values
  when the log likelihood is not included are: $\Delta\chi^2\lt59$ and
  $\Delta\chi^2\lt70$ for 70\% and 90\% confidence regions,
  respectively.

The magnitude of these differences is quite surprising.  Therefore we
perform several additional tests to understand them better.  First, we
verify that the measured differences $\Delta G$ are actually
significant with respect to fluctuations caused by modelling noise or
measurements uncertainties.  Indeed, it is natural to speculate that
the numerical noise in the procedure of the weight-adjustment may
cause fluctuations of $G$ over time. However, these
  fluctuations are comparatively small; on average they are $\sim8$ in
  G ($\sim 17$ in $\chi^2$), i.e. a factor of three smaller than the
  typical values of $\Delta G$ (and $\Delta\chi^2$).

Then, in order to quantify the relevance of the observational
measurement uncertainties, we generate a Monte Carlo chain of one
hundred realizations of the data of the NGC~4494-like mock galaxy of
Section~\ref{ssec:pseudo}.  Every realization is constructed by
drawing random Gaussian-distributed values for all kinematic data
points, such that the mean is as predicted for the mock galaxy, and
the variance corresponds to the observational errors \citep[see
\eg][]{vandermarel98,cretton00,thomas05}.  Then, we model each
realization with NMAGIC, assuming the true gravitational potential of
the target galaxy. Based on modeling $100$ realizations of
  the observational data, we conclude that random variations in the
  data according to the observational errors correspond to
  fluctuations of $\sim16$ in the merit function $G$ (and $\sim 32$ in
  $\chi^2$) within $1\sigma$.  Instead, the experiments above showed
  that if the potential is not known in advance, NMAGIC models match
  the true potential of the dark matter halo within $\Delta G\lt26$ at
  $1\sigma$ level. Hence, the $\Delta\chi^2$ difference derived from
  only fitting data sets of the true model \citep[\eg][]{thomas05}
  appears to be an underestimate.

%%%%%%%%%%%%%%%%%%%%%%%%%%%%%%%%%%%%%%%%%%%%%%%%%%%%%%%%%%%%%%%%%%%%%%%%%%%%%%%%%%%%%%%%%%%
\begin{figure}
\begin{center}
\includegraphics[scale=0.5]{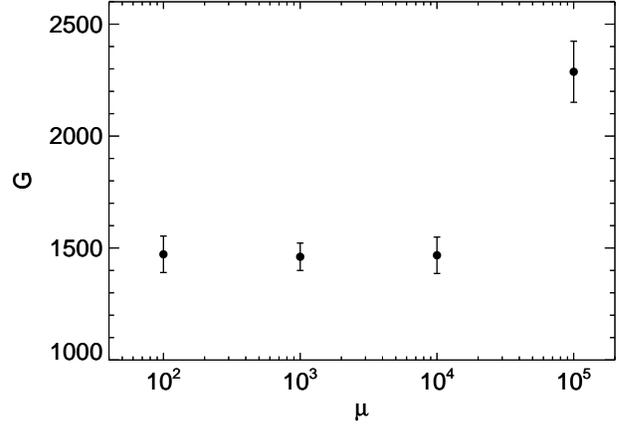}
\caption[]{The merit function $G=\chi^2/2-\mathcal{L}$ for
    increasing values of the regularization parameter $\mu$.  {\it
      Dots} represent the mean over 100 model fits to Monte Carlo
    realizations of the kinematic data of the NGC~4494-like galaxy;
    {\it error bars} represent the rms deviation from the mean
    multiplied by a factor 5.}
\label{fig:delta_mu}
\end{center}
\end{figure}
%%%%%%%%%%%%%%%%%%%%%%%%%%%%%%%%%%%%%%%%%%%%%%%%%%%%%%%%%%%%%%%%%%%%%%%%%%%%%%%%%%%%%%%%%%%

As a further check, we verify whether the minimum value of $G$ and/or
the magnitude of $\Delta G$ are influenced by the degree of
regularization employed in the NMAGIC fits.  In the framework of
Schwarzschild modelling, \citet{verolme02} showed that smoothing
constraints do not affect the shape of $\chi^2$-contours
significantly, and leave the overall best-fitting parameters
unchanged.  \citet{thomas05} constructed Schwarzschild models for
different values of the smoothing parameter, and computed the
$\Delta\chi^2$ due to $N_{\rm mock}=60$ different realizations of the
data sets as a function of the smoothing parameter.  They found that
the magnitude of $\Delta\chi^2$ increases for increasing
regularization (see their Fig.~6).  We repeat their experiment
modelling $N_{\rm mock}=100$ realizations of the data of the
NGC~4494-like mock galaxy for different values of the regularization
parameter $\mu$, and show our results in Fig.~\ref{fig:delta_mu}.  The
error bars represent the $1\sigma$-variation $\Delta G$ as a function
of $\mu$ (note that they are multiplied by a factor 5), and
show no indication that such $\Delta G$ is influenced by the
regularization, up until and including the optimal regularization
parameter used in our modeling ($\mu=10^4$).  An increase of the
fluctuations of $G$ due to regularization is only seen for
oversmoothed models ($\mu=10^5$, see Section~\ref{ssec:mu}).

The Monte Carlo experiments also reveal some slight biases which are
intrinsic to our diagnostics of the best-fitting model.  In
particular, we find that the minimum $\chi^2$ values are achieved on
average for slightly more massive halos than that of the target
galaxy, as previously noted in \citet{gerhard98}, while the maximum
$\mathcal L$ values are typically achieved for halos which are
slightly more diffuse than the true (known) halo.  The combination of
$\chi^2$ and $\mathcal L$ in the merit function $G$ in
equation~(\ref{eq:G}) is less biased, since it strikes a compromise
between these two opposite trends.

\subsection{Summary}

To summarize, we investigated how well the dark halo parameters can be
constrained from data which have the spatial coverage and error bars
of the current observational data for NGC~4494.  These data do not
suffice to determine uniquely both the scale radius and the circular
velocity of the halo specified in equation~(\ref{eq:log_halo}), and
different combinations of ($r_0$,$v_0$,$\Upsilon$) that yield similar
values of the total enclosed mass are allowed.  The enclosed mass
within $2R_{\rm e}$ can be determined to within 10\%, and the dark
matter fraction $f_{\rm DM}(3R_{\rm e})$ to within $\pm0.1$.  For the
family of logarithmic dark matter halos, the total circular velocity
can be determined to $\pm20{\rm km s}^{-1}$.

Via Monte Carlo experiments, we computed confidence levels for
parameter estimation. We conclude that, for NMAGIC models of
  the observational data at hand, the usual 1$\sigma$ (70\%) level
  corresponds to a value of $\Delta G\sim26$, and of
  $\Delta\chi^2\sim59$ (and $\Delta G\sim32$, $\Delta\chi^2\sim70$ at
  90\% confidence level). Using these statistical results, we can
now derive the uncertainties on the determination in the halo
parameters for NGC~4494.

\section{Dynamical models of NGC~4494}
\label{sec:models}

We now use NMAGIC to construct axisymmetric dynamical models for
NGC~4494 fitting all the photometric and kinematic data described in
Section~\ref{sec:data}.  Three different inclinations for the stellar
distribution are considered, $i=90\degrees,70\degrees,45\degrees$, for
which we carried out the deprojection of the photometric data (see
Section~\ref{ssec:photdata}).  The latter value is close to the
minimum inclination allowed by the observed flattening of NGC~4494
\citepalias{napolitano09,foster11}.  For each inclination, we explore
a sequence of gravitational potentials, including the self-consistent
case with constant $M/L$ and various quasi-isothermal dark matter
halos.

\begin{figure}
\centering
\includegraphics[scale=0.5]{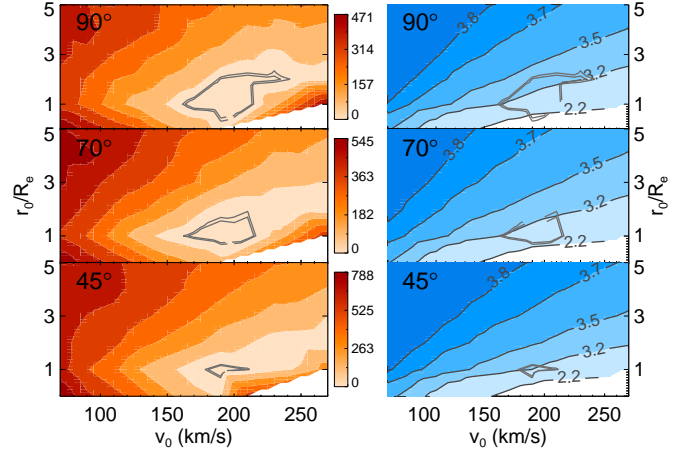}
\caption[]{Results of the NMAGIC dynamical models of NGC~4494
for the range of explored halo parameters ($r_0$,$v_0$). 
{\it From top to bottom}: $i=90\degrees,70\degrees,45\degrees$.
The {\it grey contours} correspond to $70\%$ and $90\%$ confidence levels,
as determined in Section~\ref{ssec:mc}.
{\it Left column}: $\Delta G$ of NMAGIC models relative to the best-fitting model 
(separate colour bar for each inclination). The bottom-right corner is a region
in which no good models for the data could be found.
{\it Right column}: V-band mass-to-light $\Upsilon_{\rm V}$ of the final models. 
}
\label{fig:4494_models}
\end{figure}

The main results of our suite of dynamical models are presented in
Fig.~\ref{fig:4494_models}.  In analogy with the analysis of the
previous Section, we plot the $\Delta G$ of each NMAGIC model relative
to the best-fitting model, defined as that with minimum $G$, for each
inclination $i=90\degrees,70\degrees,45\degrees$. For reference,
  the best-fitting model for $i=90\degrees$ has $\chi^2/J=0.41$, log
  likelihood $\mathcal L=-793.3$, and $G=1449.2$. In the same
figure, we also overplot the $70\%$ and $90\%$ confidence levels
determined from the Monte Carlo experiments of Section~\ref{ssec:mc}.
In the right panel, we show the final mass-to-light ratios of our
dynamical models.

\begin{figure}
\begin{center}
\includegraphics[width=10.5cm,height=11cm,trim=50 0 0 0]{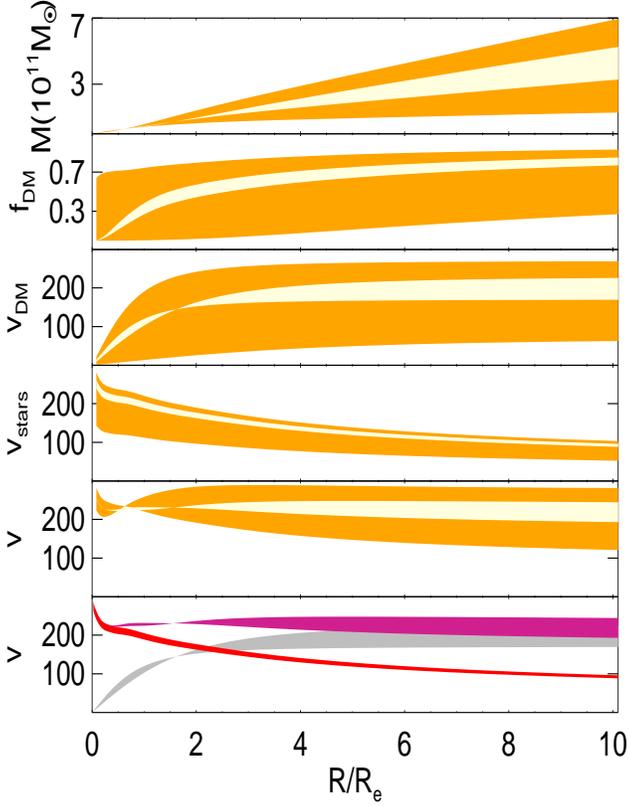}
\end{center}
\caption[]{{\it Top five panels}: as a function of radius, the
    total mass, the dark matter fraction $f_{\rm DM}=M_{\rm{DM}}/M$
    within R, and the circular velocity curves corresponding to dark
    matter, stars, and total potential. The {\it orange shaded region}
    shows all edge-on NMAGIC models investigated, the {\it yellow
      shaded region} shows the 70\% confidence region (see
    Section~\ref{ssec:mc}). {\it Bottom panel}: the contributions of
    stars ({\it red}) and dark matter ({\it grey}) to the total
    circular velocity curve ({\it violet}) for the edge-on models in
    the 70\% confidence band.}  
\label{fig:4494_vc}
\end{figure}

The shape of the $\Delta G$-contours is regular, and the range of dark
matter halos consistent with the data within the confidence levels has
circular velocity in [$160$-$240$] ${\rm km s}^{-1}$ and scale radius
$\sim1$-$2 R_{\rm e}$. The stellar mass-to-light ratio is
  $\Upsilon_V\sim3$.  The main characteristics of the mass
distribution of the explored models, and of the preferred models, are
plotted in Fig.~\ref{fig:4494_vc}.  The total circular velocity curve
of the good models is approximately flat (``isothermal'') outside
$1R_{\rm e}$, with $v_{\rm c}(3R_{\rm e})\sim220{\rm km s}^{-1}$.
Less massive halos, as well as models with constant $M/L$, are not
consistent with the data.  

It is apparent from Fig.~\ref{fig:4494_models} that these main
results, and the topology of the contours of $G$, are independent of
the adopted inclination of the stellar distribution of NGC~4494. In
general, we find that edge-on models ($i=90\degrees$) are preferred,
in the sense of a lower value of $G$, a lower value of $\chi^2$, and a
higher value of $\mathcal{L}$, respectively.  A discussion of this
issue is deferred to Section~\ref{ssec:inc}, whereas we now describe
the model fits to the data, and the orbital structure of the dynamical
models.

\subsection{Model fits to the observational data}
\label{ssec:fits}
For a wide range of dark halos, NMAGIC finds very good fits to the
observational data in terms of $\chi^2$ values. In
  particular, for all the assumed values of the inclinations, the
  models embedded in the dark matter halos that that we tried
  converged to $\chi^2/J$ values per data point less than 1 (except
  for extreme values of the parameters, corresponding to the
  bottom-right corner of Fig.~\ref{fig:4494_models}).  The models with
  constant $M/L$ achieve the poorest fit, with a value of reduced
  $\chi^2/J=0.60$ ($90\degrees$), $\chi^2/J=0.94$ ($70\degrees$), and
  $\chi^2/J=1.15$ ($45\degrees$).

\begin{figure}
\centering
\includegraphics[width=5.0cm,angle=-90]{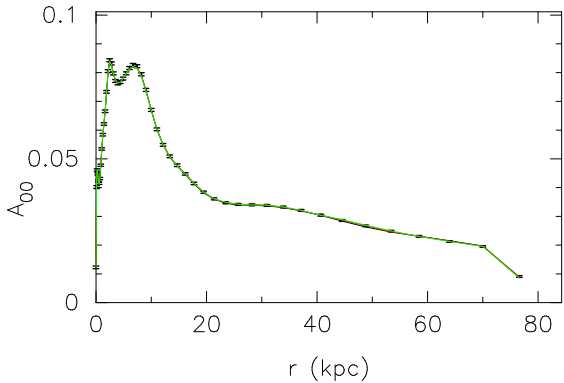}
\caption{NMAGIC fits to the differential stellar mass distribution
  ($A_{00}$ coefficient) for an inclination of 90$\degrees$.  The {\it
    black} line is for the best-fitting model ($r_0=1R_{\rm e}$,
  $v_0=190{\rm km s}^{-1}$), the {\it red} line is for a model
  residing in a more diffuse halo ($r_0=5R_{\rm e}$, $v_0=90{\rm km
    s}^{-1}$), and the {\it green} line represents a model obtained
  for a massive halo ($r_0=1R_{\rm e}$,$v_0=250{\rm km s}^{-1}$).  The
  three model curves are nearly identical even though the last two
  models are outside the $90\%$ confidence region. }
\label{fig:4494_alms}
\end{figure}

The fits to the photometric constraints are generally excellent, and
they are visually indistinguishable for most potentials, and for the
different inclinations.  Models compatible with the data at the 90\%
level achieve these fits with a maximum value of
$\chi^2/J_{A_{lm}}=0.43,0.48,0.88$, for
$i=90\degrees,70\degrees,45\degrees$, respectively.
Fig.~\ref{fig:4494_alms} shows the very good fit to the first moment
of the $A_{lm}$, i.e. the differential stellar mass distribution, for
the preferred inclination of 90$\degrees$.  Three fiducial halo models
are shown: the best-fitting model, and two models which fit the data
with $\chi^2<1$, but lie outside the 90\% confidence band, having a
more massive and a more diffuse dark halo than the best-fitting
models.

\begin{figure}
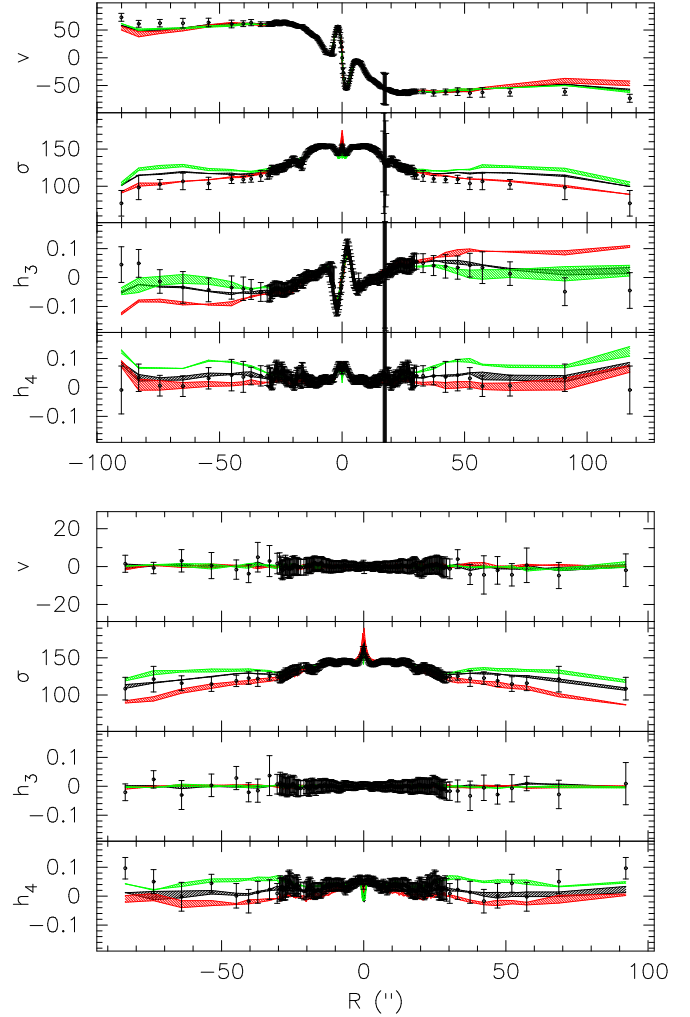

\includegraphics[scale=0.41,angle=-90.0]{4494_slit_maj.ps}
\vskip.5cm
\includegraphics[scale=0.41,angle=-90.0]{4494_slit_min.ps}
\caption[]{NMAGIC fits to the long-slit kinematic data ({\it black
    dots}) of NGC~4494 along the major ({\it top}) and minor ({\it
    bottom}) axis.  The {\it black shaded region} shows the
  best-fitting edge-on models within the 90\% confidence range.  The
  {\it red and green shaded regions} show the spread of two fiducial
  models for the explored inclinations; {\it red} is for the model
  residing in a lighter halo ($r_0=5R_{\rm e}$, $v_0=90{\rm km
    s}^{-1}$), and {\it green} is for the model obtained in a massive
  halo ($r_0=1R_{\rm e}$, $v_0=250{\rm km s}^{-1}$). Both models are
  well outside the confidence range.}
\label{fig:4494_slit}
\end{figure}

The fits to the long-slit kinematics are also generally good.  Models
within the 90\% confidence range achieve a maximum value of
$\chi^2/J_{\rm Sl}=0.36,0.41,0.35$, for
$i=90\degrees,70\degrees,45\degrees$, respectively.  For the fiducial
models, these fits are shown in Fig.~\ref{fig:4494_slit}, where it is
apparent that the central feature of the kinematically decoupled core
in $v$ and in $h_3$ is reproduced well by our particle models, and
that the models have some difficulty in matching the detailed
long-slit kinematics along the outer major axis.  In particular, the
best-fitting models have slightly lower values of $v$ and slightly
higher values of $\sigma$.  To a great extent, these systematic
deviations can be traced to the compromise that the models must find
(see below) between the long-slit and the slitlets kinematic data,
which extend to larger radii.  The fiducial models show that more
massive/diffuse halos result in higher/lower values of $\sigma$ and
$h_4$ along both axes.  Finally, the shaded regions in
Fig.~\ref{fig:4494_slit} show the range of models allowed by the
different inclinations (for the more massive and the more diffuse
fiducial models), and by the 90\% confidence region (for the
best-fitting model).

\begin{figure}
\includegraphics[scale=0.48]{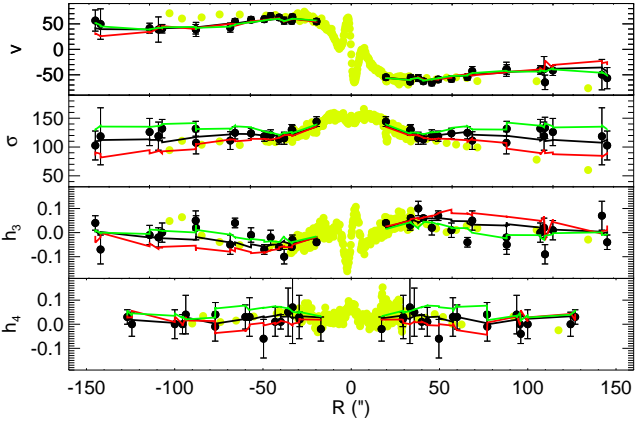}
\caption[]{NMAGIC fits to the slitlets kinematics near the major axis
  of NGC~4494 ({\it black dots} with error bars).  For comparison,
  {\it light-green dots} represent the original major-axis
  slit data.  The model points are averages over the same slit cells
  as the data, and are connected by straight line segments.  Different
  colours represent the fiducial models: {\it black} is for the
  best-fitting model ($r_0=1R_{\rm e}$, $v_0=190{\rm km s}^{-1}$),
  {\it red} is for the model residing in a lighter halo ($r_0=5R_{\rm
    e}$, $v_0=90{\rm km s}^{-1}$), and {\it green} is for the model
  obtained in a massive halo ($r_0=1R_{\rm e}$, $v_0=250{\rm km
    s}^{-1}$).  For simplicity, only the preferred edge-on models are
  shown.  }
\label{fig:4494_lets}
\end{figure}

The fits to the newer slitlets kinematics are achieved with maximum
$\chi^2/J_{\rm Lets}=1.,1.18,1.27$ within the 90\% confidence level,
for the three inclinations of $90\degrees,70\degrees,45\degrees$,
respectively.  Fig.~\ref{fig:4494_lets} shows the fits of the
preferred edge-on particle models to the slitlets kinematics for the
fiducial potentials.  The trends with halo mass described before are
still clearly visible.  Moreover, it can be seen that the best-fitting
edge-on model, corresponding to the black line, has to find a
compromise between the kinematic data in slit and slitlets in the
outer region.  This results in a lower magnitude for the velocity, and
in a higher velocity dispersion, which explains the deviations from
the long-slit kinematics in the outermost data points.

\begin{figure*}
\centering
\includegraphics[width=11cm,angle=-90]{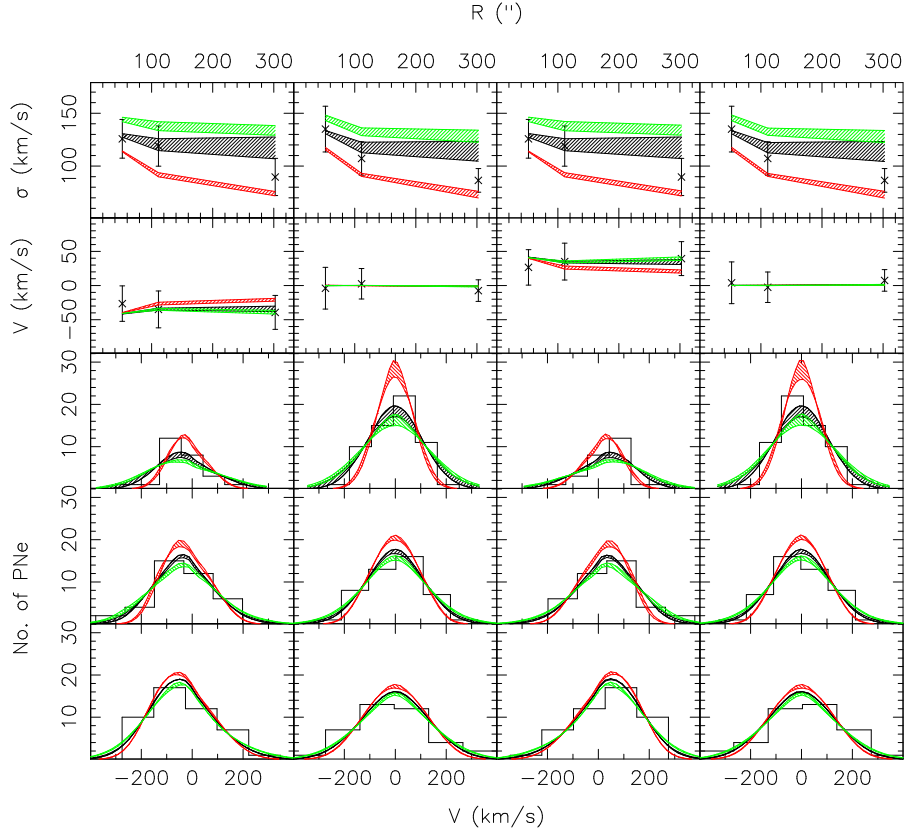}
\caption{NMAGIC fits to the PNe velocities for all the considered
  inclinations.  {\it Top two rows}: mean velocity dispersion and
  velocity profiles in the angular segments centered on 0$\degrees$
  (major axis), 90$\degrees$, 180$\degrees$, 270$\degrees$, from left
  to right. {\it Bottom three rows}: LOSVD in the same angular
  segments at radii of $52''$, $110''$, and $300''$, going upwards.
  Colours of the {\it shaded regions} are as in
  Fig.~\ref{fig:4494_slit}.  }
\label{fig:4494_pne}
\end{figure*}

Finally, Fig.~\ref{fig:4494_pne} shows a comparison of the final
NMAGIC models with the PN kinematic data, for all the inclinations and
for the fiducial potentials.  The velocity, velocity dispersion, and
LOSVD are plotted in the angular segments used for the modelling (see
Section~\ref{sec:pne}).  The axisymmetric nature of our models is
apparent in the reflection symmetry of the kinematics in diagonally
opposite segments.  While the mean velocity profiles are well fit by
all models shown, the low (high) dark matter models are systematically
low (high) in the PN dispersion plot, and show systematic deviations
from the data histograms.  The best-fitting models, instead, provide a
good match to all the PN data.

\subsection{The internal kinematics of NGC~4494}
\label{ssec:beta}
\begin{figure}
\begin{center}
\includegraphics[angle=-90.0,scale=0.37]{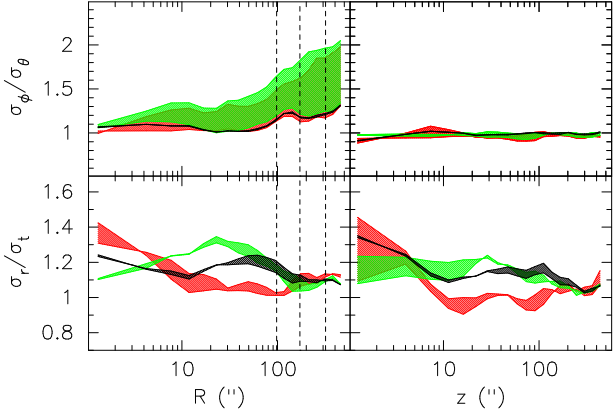}
\caption[]{Internal kinematics of the final NMAGIC models for
  NGC~4494.  The ratio of the azimuthal to meridional velocity
  dispersion ({\it top}) and of the radial to tangential velocity
  dispersion $\sigma_{\rm t} = \sqrt{(\sigma_{\theta}^2+
    \sigma_{\phi}^2)/2}$ ({\it bottom}) are plotted as a function of
  major axis $R$ in the equatorial plane ({\it left column}), and of
  minor axis $z$ in the meridional plane ({\it right column}).
  Colours of the {\it shaded regions} are as for
  Figs.~\ref{fig:4494_slit},~\ref{fig:4494_pne}.  The vertical {\it
    dashed lines} show the extent of the kinematic data (long-slit,
  slitlets, and PNe, going outwards).  }
\label{fig:4494_int}
\end{center}
\end{figure}

We now look at the intrinsic kinematics of the NGC~4494 models, in
order to learn about the orbital anisotropy of this galaxy.
Fig.~\ref{fig:4494_int} shows the internal kinematics for the range of
explored inclinations.

%%%%%%%%%%%%%%%%%%%%%%%%%%%%%%%%%%%%%%%%%%%%%%%%%%%%%%%%%%%%%%%%%%%%%%%%%%%%%%%%%%%%%%%%%%%
%%%%%%%%%%%%%%%%%%%%%%%%%%%%%%%%%%%%%%%%%%%%%%%%%%%%%%%%%%%%%%%%%%%%%%%%%%%%%%%%%%%%%%%%%%%

As expected, in the radial region well constrained by the kinematic data
the ratio of the radial to tangential velocity dispersions is larger
for models embedded in a massive halo than for low dark matter models
\citep[see \eg][]{binmam82,gerhard93}.

On average, the ratio of the radial to tangential velocity dispersions
along $R$ in the best-fitting models is $\sim1.2$, i.e. the underlying
orbital distribution is slightly radially biased.  As discussed
further in Section~\ref{sec:discuss}, the degree of radial anisotropy
of NGC~4494 is milder than what was previously found for the two other
intermediate-luminosity ellipticals NGC~4697 \citepalias{dl08} and
NGC~3379 \citepalias{dl09}.

For all the explored potentials, the two components of the tangential
velocity dispersion are similar along the minor axis, as required for
an axisymmetric system, whereas the azimuthal dispersions are higher
in the equatorial plane, suggesting together with the radial
anisotropy that this (rotating) elliptical may be flattened by
meridional anisotropy \citep[\eg][]{dehger93,thomas09b,das11}.
Variations due to the assumed inclination are always larger in the
ratio between the azimuthal and meridional velocity dispersions along
$R$.

Finally, at larger radii, and close to the outermost data points
(marked by the vertical dashed lines), the orbital structure is nearly
isotropic along both $R$ and $z$.  As we have shown in
\citetalias{mg11}, in those external regions, which are unconstrained
by the observational data, a bias towards the dynamical structure of
the initial (isotropic, in this case) particle model cannot be
avoided.

\subsection{The inclination of NGC~4494}
\label{ssec:inc}
From our NMAGIC models of NGC~4494, we would conclude that the
observational data prefer an inclination of 90$\degrees$ out of the
three inclinations that we explored.  Indeed, edge-on models provide
better fits to the data than models with lower inclinations, in terms
of the merit function $G$, as well as $\chi^2$ and $\mathcal{L}$.
In particular, we find a $\Delta G\sim144$ (78) between the
  best-fitting edge-on model and the best-fitting model obtained for
  $45\degrees$ $(70\degrees)$, which is highly significant when
  compared to the typical differences $\Delta G$ among different
  potentials for a given inclination. Visually, it is hard to
distinguish the fit to the observables for the $90\degrees$ and the
$70\degrees$ inclinations, whereas the more face-on $45\degrees$
models are characterized by a systematically low velocity and velocity
dispersion with respect to the observational data.

Should this preference for edge-on inclinations really be believed?
Up to now, we have assumed a spherical dark halo for all the
considered inclinations, and only the stellar distribution was
axisymmetric and varied according to the inclination.  We can relax
the assumption of sphericity of the dark matter halo, and investigate
dark matter halos with similar flattening as for the stars.  To this
end, we consider the flattened logarithmic potential
\begin{equation} 
 \phi_D(R,z)=\frac{v_0^2}{2}\ln\left(r_0^2+R^2+\frac{z^2}{q_\phi^2}\right)
\label{eq:log_halo_q}
\end{equation}
\citep{bt08}. Then, we use the approximate relation
$1-q_\phi\sim(1-q_\rho)/3$ to derive the flattening of the potential
$q_\phi$ from that of the density distribution $q_\rho$, which we
infer from the apparent flattening $q$ of the isophotes via
\begin{equation} 
 q^2=\cos^2{i}+q_\rho^2\sin^2{i}.
\end{equation}
For an average flattening of the isophotes $q=0.82$, the intrinsic
flattening of the density distribution $q_\rho=0.82,0.79,0.59$ for
$i=90\degrees,70\degrees,45\degrees$, respectively, and the
corresponding $q_\phi=0.94,0.93,0.86$.

We have explored a few flattened dark halos with these parameters and
the same spherically-averaged mass distribution as the best-fitting
models for $i=70\degrees$ and $45\degrees$. Although some of them
provide slightly better fits to the observational data, our main
result does not change, and edge-on models are still preferred by far.

The obvious next question is whether there could be some bias in the
modelling technique.  This explanation was proposed by
\citet{thomas07_2}, who found that all their best-fitting
Schwarzschild models of $N$-body merger remnants have $i=90\degrees$,
and argued that edge-on models necessarily have a smaller $\chi^2$
than face-on models, due to their greater freedom in the adjustment of
prograde and retrograde orbits to fit rotation and axisymmetric
deviations from a Gaussian LOSVD.

We used our NGC~4494-like galaxy (see Section~\ref{ssec:pseudo}) to
test whether particle models exhibit a bias analogous to that of
orbit-based models.  First, we constructed a series of NGC~4494-like
galaxies assuming different inclinations $i_{\rm
  G}=90\degrees,70\degrees,45\degrees$ for the stellar
distribution. Then, we modelled the observational data of these mock
galaxies with NMAGIC for different inclinations $i_{\rm M}$.  The
results of these experiments are summarized in Table~\ref{tab:inc},
and confirm the findings and the argument of \citet{thomas07_2}: on
average, the values of $\chi^2$, and of the merit function $G$,
increase for lower values of the inclination.  Table~\ref{tab:inc}
reveals that edge-on models generally provide better fits.  This is
true both in the case in which the model has the true known
inclination of the target galaxy, and in the case in which the model
assumes a wrong value of the inclination.  For reasons that are
currently unclear to us, the likelihood favours instead more face-on
models.

%%%%%%%%%%%%%%%%%%%%%%%%%%%%%%%%%%%%%%%%%%%%%%%%%%%%%%%%%%%%%%%%%%%%%%%%%%%%%%%%%%%%%%%%%%%
\begin{table}
\centering
\begin{tabular}{c c c c c}
\hline
$i_{\rm G}$ & $i_{\rm M}$ & $\chi^2$&$G$& -$\mathcal{L}$\\ 
\hline
$90\degrees$ & $90\degrees $& 0.24 & 1148.74 & 769.93 \\
$70\degrees$ & $70\degrees $& 0.27 & 1195.40 & 767.99 \\
$45\degrees$ & $45\degrees $& 0.35 & 1313.38 & 754.93 \\
$90\degrees$ & $70\degrees $& 0.26 & 1178.91 & 769.84 \\
$90\degrees$ & $45\degrees $& 0.38 & 1371.26 & 770.90 \\
$45\degrees$ & $70\degrees $& 0.30 & 1228.82 & 756.56 \\
$45\degrees$ & $90\degrees $& 0.29 & 1218.26 & 756.86\\
\hline
 \end{tabular}
 \caption{Results of NMAGIC fits to an NGC~4494-like galaxy 
     whose stellar distribution has a known inclination $i_{\rm G}$,
     assuming an inclination $i_{\rm M}$ in the modelling.
     $\chi^2$ values are normalized by the number of observables
     $J=3186$.}
\label{tab:inc}
\end{table}
%%%%%%%%%%%%%%%%%%%%%%%%%%%%%%%%%%%%%%%%%%%%%%%%%%%%%%%%%%%%%%%%%%%%%%%%%%%%%%%%%%%%%%%%%%%

This test does not only confirm the existence of a bias
\citep[e.g.,][]{thomas07_2}, but also enables us to quantify its
effect on the dynamical models of NGC~4494. The experiments
  show that, when modelling a NGC~4494-like galaxy which has a true
  inclination $i_{\rm G}=45\degrees$, a $\Delta G\sim95$
  between the edge-on model and the more inclined model could be
  associated with the ``edge-on bias". However, our NGC~4494 runs
display a significantly larger difference $\Delta G\sim144$ 
between the edge-on and the $i_{\rm M}=45\degrees$ models.  
Thus, combining these results
suggests that NGC 4494 is truly close to edge-on. If so, this might
provide also a simple explanation for the slightly larger rotation
velocities measured in the major axis slit compared to the surrounding
slitlets (see Fig.~\ref{fig:4494_lets}). This could be the kinematic
signature of a faint disk.

\section{Discussion}
\label{sec:discuss}
In this Section we discuss ({\it i}) the results on parameter
estimation of dark matter halos with NMAGIC, ({\it ii}) our dynamical
models for NGC~4494 in comparison with previous work, and ({\it iii})
the dark matter distribution and orbital structure of NGC~4494 in the
wider context of intermediate-luminosity elliptical galaxies with
steeply falling, 'quasi-Keplerian' velocity dispersion profiles.

\subsection{Confidence limits for parameter estimation with NMAGIC}

In Section~\ref{sec:test} we used Monte Carlo simulations of a
NGC~4494-like mock galaxy in different dark matter haloes to assess
the confidence limits for the halo parameters that provide
statistically valid models for the real NGC 4494.  The data set for
the mock galaxy closely resembled that of NGC~4494.  We estimate the
differences in the merit function $\Delta G$ corresponding to
confidence levels enclosing a certain probability of finding the true
values of the parameters \citep[\eg][]{press92,thomas05}.
Specifically, we constructed and modelled $10$ NGC~4494-like galaxies
in different dark matter halos, and estimated the distribution of
$\Delta G$ values between the best-fitting model and the model with
the true halo parameters for our data and modelling set-up.

These experiments showed that our NMAGIC dynamical models match
  the true potential of the dark matter halo within $\Delta G\lt26$
  ($\Delta\chi^2\lt59$) about 70\% of the time, and within $\Delta
  G\lt32$ ($\Delta\chi^2\lt70$) about 90\% of the time. These values
also correspond approximately to the fluctuations found within the
confidence boundaries.  The numerical noise in $G$ or $\chi^2$ caused
by the adjustment of particle weights explains only a minor fraction
of these $\Delta G$ or $\Delta\chi^2$ fluctuations.  Fluctuations
induced by varying the data within their error bars are much more
significant compared to the measured $\Delta G$ or $\Delta\chi^2$
values, as already noted by \citet{thomas05} in the context of
Schwarzschild's method.  Our experiments show, however, that some
additional uncertainty must be related to the freedom associated with
comparing different trial potentials. They also indicate that these
results are unaffected by the strength of regularization as long as
the model is not oversmoothed.

The relative differences $\Delta G$ (and $\Delta\chi^2$) found in our
simulations are substantially larger than what is often assumed, and
reveal that the $\Delta\chi^2=(1, 2.3, ..)$ criterion for
  $\nu=(1, 2, ..)$ free model parameters used in many dynamical
studies in the literature (see Introduction) is inappropriate for our
NMAGIC particle models.

The $\Delta\chi^2(\nu)$ approach in dynamical modelling is based
  on $\chi^2$-statistics and assumes Gaussian errors, linear
  dependence on the model parameters, and that the number of degrees
  of freedom is positive \citep[\eg][]{press92}. Then $\chi^2$ can be
  marginalized over all degrees of freedom other than the number of
  free parameters in the mass model. This implies the assumption that
  much of the freedom in the fitted weights (or orbits) is used by the
  regularization in constraining the distribution function from the
  data in the given trial potential.  Because the number of particle
  weights is much larger than the number of data points, and given the
  indications that the simulations results are insensitive to the
  regularization parameter (Section~\ref{ssec:mc}), this assumption
  could be incorrect.

The $\Delta\chi^2$ values found in our experiments appear to be
  larger than in typical Schwarzschild applications.  One exception
  appears to be the work of \citet{vandenbosch+vandeven09} who
  determined confidence limits on the modelling parameters based on
  the expected standard deviation of $\chi^2$ itself,
  $s(\chi^2)=\sqrt{2(J-M)}$, where $J$ is the number of observational
  constraints and $M=4$ the number of their free model
  parameters. However, the fact that the standard deviation of
  $\chi^2$ {\sl is} $s(\chi^2)$ does not invalidate the standard
  $\Delta\chi^2(\nu)$ method when the underlying assumptions
  \citep[\eg][]{press92} are met. Thus, while the $\Delta\chi^2$ values
  found in our Monte Carlo simulations do have the same order of
  magnitude as $s(\chi^2)$, it is not clear whether this is
  significant until more simulations in different modelling contexts
  have been analyzed.

The experience in dynamical modelling with made-to-measure particle
models is still limited, and it is possible that there are some
  aspects of these methods that influence parameter estimation in
different ways than, {\it e.g.}, Schwarzschild models.  However, given
the wide-spread use of dynamical modelling for measuring, {\it e.g.},
dark halo parameters and black hole masses in galaxies, it is
important to test the statistical premises of this work more
thoroughly.  Monte Carlo simulations like those we have performed may
be the best way to tackle these issues, and determine the appropriate
$\Delta\chi^2$ values for estimating confidence limits for a given
observational and modelling set-up.

\subsection{Dynamical models for NGC~4494: comparison with the literature}

The elliptical galaxy NGC~4494 has previously attracted the attention
of several dynamical studies
\citep[\eg][]{vandermarel91,kron00,mago01,rom03,rod11,napolitano09,lackner10}.

Our NMAGIC models improve on the models explored so far in some
important aspects.  First, we considered as many observational data as
currently available, i.e.  at large radii we had available both PNe
velocities \citepalias{napolitano09} and galaxy spectra in slitlets
\citepalias{foster11}.  The newer slitlets data show a milder drop of
the velocity dispersion than what it is suggested by the long-slit
data \citepalias{foster11}, and this is likely to account for the
larger enclosed mass beyond $\sim2R_{\rm e}$ in our results compared
to the Jeans models of \citetalias{napolitano09}.  Second, because of
the greater constraining power of the new data compared to the PNe, we
sampled the dark halo parameter space much more finely than before,
and for different values of the inclination
($i=45\degrees,70\degrees,90\degrees$).  Thirdly, we performed a
thorough analysis of the confidence levels at which these parameters
can be estimated with our models, given the data at hand.  Finally,
contrary to most previous studies that considered spherical models
\citep[but see][]{rod11}, our NMAGIC models are axisymmetric, and we
also explored the possibility of flattened dark halos.  While the most
robust results should eventually be derived using triaxial models, it
has been shown that relaxing the spherical assumption hardly
influences the recovered halo mass \citepalias[see \eg][]{dl09}.

Although NGC~4494 was among the three galaxies described as ``naked''
in \citet{rom03}, and has an unusually low dark matter fraction in the
analysis of \citet{deason12}, our dynamical models show that a dark
matter halo is required.  In this respect, they agree with
\citet{rod11}, although the small sequence of explored models did not
allow them to put robust constraints on the halo mass, and also with
\citetalias{napolitano09}.  \citetalias{napolitano09} investigated a
family of multi-component kurtosis-based Jeans models. Their
  favoured logarithmic dark halo had $r_0=4R_{\rm e}$ and $v_0=150{\rm
    km s}^{-1}$.  Our best-fitting halo models are more massive than
  what they obtained (see Section~\ref{sec:models}), although probably
  consistent with their errors.  These differences are probably mostly
  due to the additional slitlets constraints that we included, and the
  better use of the higher-order kinematic moments in the NMAGIC
  modelling.

The dark matter fraction of our best-fitting NMAGIC models $f_{\rm
  DM}(<5R_{\rm e})=0.6\pm0.1$ is therefore higher than what was
reported by \citetalias{napolitano09} ($0.2$-$0.5$).  Interestingly,
the new value agrees very well with the stellar population predictions
computed by \citet{deason12} assuming a Chabrier initial mass function
(see their Fig.~7).  Their spherical distribution function models of
the PNe velocities, on the other hand, indicated a low dark matter
fraction $f_{\rm DM}(<5R_{\rm e})=0.32\pm0.12$.

Alternative estimates of the mass of NGC~4494, \eg from the X-ray
emission of hot gas, would be highly desirable.  From the compact
X-ray gas emission, \citet{fukazawa06} estimated
$\Upsilon_B=6.2\pm1.9$ inside $1R_{\rm e}$, consistent with the Jeans
modelling of \citetalias{napolitano09} but higher than our NMAGIC
results.  However, the existence of X-ray emitting gas around this
galaxy has been questioned \citep{osullivan04,diehl07}, and the
validity of hydrostatic equilibrium may also be dubious \citep{cp96}.

Additional kinematic constraints at large radii could be obtained from
different tracers of the mass distribution, such as GCs.  However,
from the analysis of the spatial and kinematic distributions of blue
and red GCs around NGC~4494, which do not follow that of the stars
\citepalias{foster11}, it is likely that PNe and GCs are in distinct
dynamical equilibria in the same gravitational potential \citep[see
\eg][]{payel2}.

Our best-fitting NMAGIC models have a mass-to-light
$\Upsilon_B=3.71\pm0.15$, obtained converting from $\Upsilon_{\rm V}$
using the de-reddened colour from \citet{goudfrooij94}.  This value of
the stellar mass-to-light ratio is lower than that obtained from the
Jeans models of \citetalias{napolitano09}, and is easier to reconcile
with independent measurements from stellar population models, which
gave $\Upsilon_B=4.3\pm0.7$ for a Kroupa initial mass function
\citepalias[see][]{napolitano09}.  In this respect, \citet{lackner10}
fitted galaxy formation model to the PNe velocities of NGC~4494, and
found that the best-fit dissipational and dissipationless models give
$\Upsilon_B=2.97$ and $\Upsilon_B=3.96$, respectively, so that a
purely dissipational formation scenario for NGC~4494 seemed to be
ruled out \citep{lackner10}.

Finally, the edge-on NMAGIC models that best reproduce the
observational data of NGC~4494 are mildly radially anisotropic, with
$\beta\sim0.4$ for the 90\% confidence range.  More inclined NMAGIC
models ($i=45\degrees$) within the same confidence range are more
radially anisotropic, i.e. $\beta\sim0.6$.  The degree of orbital
anisotropy is consistent with the previous analysis of
\citetalias{napolitano09} and \citet{deason12}.

\subsection{The wider context: intermediate-luminosity ellipticals}
\label{ssec:intermediate}
NGC~4494 belongs to the group of intermediate-luminosity elliptical
galaxies with rapidly falling velocity dispersion profile
(NGC~821, NGC~3379, NGC~4494, NGC~4697), dubbed ``naked'' by
\citet{rom03} because of the unusually low dark matter content
revealed by their favoured dynamical models.  The analysis of
\citet{coccato09} identified a larger group of early-type galaxies
(NGC~821, NGC~3377, NGC~3379, NGC~4494, NGC~4564, NGC~4697) with
strongly decreasing $v_{\rm rms}=\sqrt{\sigma^2+v^2}$ profiles.

Detailed dynamical models have now been obtained for all of the
galaxies in the original sample of \citet{rom03}, including
observational constraints from the higher order moments of the LOSVD
\citep[see the summary in \citetalias{napolitano09} and][for
NGC~821]{weijmans09,forestell10}.  On the whole, the results
suggest that these intermediate-luminosity galaxies are
inconsistent with the previous claims of little to no dark matter
halo.

We now compare the results that we derived for NGC~4494 in this work,
with the findings previously obtained by \citetalias{dl08} and
\citetalias{dl09} for NGC~4697 and NGC~3379 using the same modelling
technique.  Indeed, the photometry and kinematics of these three
galaxies are not very different. They all have intermediate values of
luminosity and stellar mass, similarly low values of the central
velocity dispersion $\sim 150-210{\rm km s}^{-1}$, and similarly
falling $v_{\rm rms}$ profiles \citep{coccato09}.

\begin{figure}
\begin{center}
\includegraphics[scale=0.45]{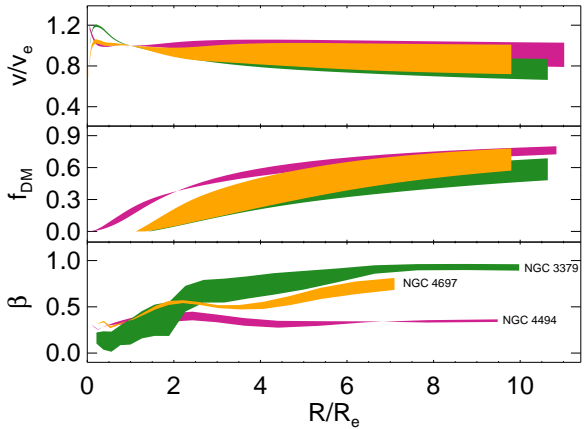}
\caption[]{{\it From top to bottom}: as a function of radius, circular
  velocity normalized by its value at $1R_{\rm e}$, dark matter
  fraction, and anisotropy parameter of the range of valid NMAGIC
  models obtained fitting the data of NGC~4494 (this work, {\it
    violet}), NGC~4697 (\citetalias{dl08}, {\it orange}), and NGC~3379
  (\citetalias{dl09}, {\it green}).  }
\label{fig:compare_int_ell}
\end{center}
\end{figure}

Fig.~\ref{fig:compare_int_ell} shows the comparison of the NMAGIC
models obtained for NGC~4697, NGC~3379, and NGC~4494.  For NGC~4494,
we plot the range of best-fitting edge-on models as determined in
Section~\ref{sec:models}.  For the other intermediate-luminosity
ellipticals, no rigorous $\Delta\chi^2$ analysis was performed, but a
range of valid models was determined based on the likelihood of PNe.

%%%%%%%%%%%%%%%%%%%%%%%%%%%%%%%%%%%%%%%%%%%%%%%%%%%%%%%%%%%%%%%%%%%%%%%%%%%%%%%%%%%%%%%%%%%
%%%%%%%%%%%%%%%%%%%%%%%%%%%%%%%%%%%%%%%%%%%%%%%%%%%%%%%%%%%%%%%%%%%%%%%%%%%%%%%%%%%%%%%%%%%

As can be seen in the bottom panel of Fig.~\ref{fig:compare_int_ell},
the orbital structure of the three intermediate luminosity ellipticals
is radially biased, in agreement with previous dynamical models of
elliptical galaxies in the literature
\citep[\eg][]{merritt97,rix97,gerhard98,vandermarel98,matger99,gebhardt00,saglia00},
with predictions of the monolithic collapse scenario
\citep{vanalbada82}, and with simulations of both binary-mergers
\citep[\eg][]{gerhard81,dekel05,thomas09b} and cosmological mergers
\citep[\eg][]{abadi06,onorbe07}, though the remnants of binary mergers
can exhibit a variety of orbital distributions \citep{naab06}.

NGC~4494 is characterized by a milder level of radial anisotropy than
the other two galaxies, consistent with the fact that its projected
velocity dispersion is higher (see Fig.~15 of \citet{coccato09}) and
its surface density profile steeper.  Recent simulations have shown
that dissipational processes in wet mergers may decrease the level of
radial anisotropy \citep[\eg][]{naab06,thomas09}, and also explain
many observed features such as counter-rotating disks, kinematically
decoupled components, and extra light at small radii
\citep[\eg][]{mihos94,springel05,cox06,jesseit07,hopkins08,hoffman10}.

All three galaxies have been classified as fast rotators
\citep{emsellem11}.  For NGC~4494, its central kinematically decoupled
core \citep{davor11}, and the small value of $\lambda_{\rm R}$ at
large radii \citep{coccato09}, suggest a complicated rotation
structure. Such transitions in $\lambda_{\rm R}$ towards large radii
\citep[for other cases, see][]{proctor09,coccato09} may be a signature
of merger events \citep[\eg][]{hoffman10}. The preference for edge-on
models found in Section~\ref{ssec:inc} also suggests that only a 
small fraction of the kinetic energy in this galaxy is in rotation.

The top panel of Fig.~\ref{fig:compare_int_ell} shows the circular
velocity curves normalized at $1R_{\rm e}$.  At large radii, the
behaviour of the circular velocity is not so different, partially due
to the assumption of logarithmic halo profiles. The greatest
discrepancies can be seen in the central regions, and are probably due
to different imprints left by baryonic processes during galaxy
formation.

This is also supported by the fact that the dark matter fractions of
these three galaxies are remarkably different (middle panel of
Fig.~\ref{fig:compare_int_ell}), while their global circular velocity
curves are more similar.  The dark matter fraction of NCG~4494 has
much higher values, particularly inside $3R_{\rm e}$, than those of
NGC~4697 and NGC~3379.  Such a high dark matter fraction could be a
consequence of a merger event (like those advocated by
\citet{proctor09} or \citetalias{foster11}), which might have
scattered dark matter into the inner regions of the galaxy
\citep{oser12,hilz12}.

\section{Conclusions}
\label{sec:conclusions}
We presented dynamical models for the intermediate luminosity
elliptical galaxy NGC~4494, fitting photometric and kinematic
observational data to investigate its mass distribution and orbital
structure.  Our extended kinematic data included the recently
available integrated light spectra in slitlets \citep{foster11} and
hundreds of planetary nebulae velocities reaching out to $\simeq 7
R_{\rm e}$ \citep{napolitano09}.  We used the $\chi^2$-made-to-measure
particle code NMAGIC to construct axisymmetric models for various
inclinations exploring a large sequence of gravitational potentials.

In parallel, we carried out a parameter estimation study,
investigating how well the characteristic parameters of dark matter
halos can be recovered via NMAGIC modelling of the available
observational data.  For this, we used Monte Carlo simulations of
NGC~4494-like mock galaxies to determine the confidence regions around
the best-fitting model. These confidence bands were then used to
discriminate the range of valid models for NGC~4494.

Our main results can be summarized as follows:
\begin{itemize}

\item Given the observational data of NGC~4494 and our NMAGIC
  modelling set-up, Monte Carlo simulations showed that the usual
  $1\sigma$ (70\%) confidence level corresponds to a relative
  difference in the merit function $\Delta G=26$ ($\Delta\chi^2=59$).
  At 90\% confidence level, $\Delta G=32$ and $\Delta\chi^2\sim70$.
  These differences are much larger than the commonly used
  $\Delta\chi^2$ values based on varying a small number of model
  parameters.

\item Under the assumption that the dark matter halo has a
    logarithmic parameterization, the best-fitting models for
  NGC~4494 determined within these confidence levels have an
  approximately flat total circular velocity curve outside
  $\sim0.5R_{\rm e}$, with $v_{\rm c}(3R_{\rm e})\sim220{\rm km
    s}^{-1}$.  The larger variation is in the dark matter circular
  velocity curve, rather than in the stellar one.  The dark matter
  fraction of the models within 70\% confidence level is about
  $0.6\pm0.1$ at $5R_{\rm e}$, and they are embedded in concentrated
  dark halos ($r_0\sim1$-$2R_{\rm e}$) with circular velocity
  $\sim[160$-$230]{\rm km s}^{-1}$.  With this large dark matter
  fraction, the stellar mass-to-light ratio is consistent with the
  value predicted by \citet{deason12} from stellar population models
  with a Chabrier IMF.  The discrepancy with the more diffuse halo
  found by the Jeans models of \citet{napolitano09} is likely due to
  the additional slitlets constraints, which suggest a milder drop of
  the velocity dispersion than shown by the long-slit kinematics.

\item The edge-on models provide the best fits to the available
  observational data, but the inferred dark halo parameters
  ($r_0$,$v_0$) do not depend sensitively on the assumed inclination.
  We explored a sequence of flattened dark matter halos, which however
  did not change our main result: edge-on models provide better fits
  to the data than models with lower inclinations.

\item The orbital anisotropy of the stars is increasingly radial from
  the center outwards, but the amount of radial anisotropy is smaller
  than what was found in similar previous works for the two other
  intermediate luminosity ellipticals with rapidly falling, quasi-Keplerian velocity dispersion profiles, NGC~3379 and
  NGC~4697.

\item Comparing the halos of all three intermediate-luminosity,
    quasi-Keplerian ellipticals modelled with the same
  made-to-measure particle technique (NGC~3379, NGC~4697 and
  NGC~4494), we conclude that they have similar global circular
  velocity curves and outer dark matter halos.  NGC~4494 shows a
  particularly high dark matter fraction inside $\sim 3R_e$, and a
  strong central concentration of baryons.  These differences are
  probably related to the detailed interplay between baryons and
    dark matter in the processes which shaped these galaxies.

\end{itemize}

\section*{ACKNOWLEDGEMENTS}
The authors thank Caroline Foster and Duncan Forbes for providing the
slitlets data before publication, Nicola Napolitano for useful
discussions, and an anonymous referee for a careful reading
  of the paper.  LM acknowledges support from and participation in
the International Max-Planck Research School on Astrophysics at the
Ludwig-Maximilians University. LC acknowledges financial support from
the European Community's Seventh Framework Programme (/FP7/2007-2013/)
under grant agreement No. 229517. IMV acknowledges support from the
DFG Priority Program SPP 1177.

\bibliographystyle{mn2e}
\bibliography{4494}

\clearpage

\end{document}